\documentclass[
 reprint,
 superscriptaddress,
 floatfix,
 amsmath, amssymb,
 aps,
pra,
]{revtex4-2}

\usepackage[utf8]{inputenc}  
\usepackage[T1]{fontenc}
\usepackage[british]{babel}
\usepackage[table]{xcolor}
\usepackage[scaled=0.86]{berasans}

\makeatletter
\newcommand{\setword}[2]{%
  \phantomsection
  #1\def\@currentlabel{\unexpanded{#1}}\label{#2}%
}
\makeatother
\usepackage{comment}
\usepackage{graphicx}
\usepackage[babel]{microtype}
\usepackage{amsmath,amssymb,amsthm,bm,amsfonts,mathrsfs,bbm}

\usepackage{xspace}
\usepackage{pgf,tikz}
\usepackage{array}
\usepackage{bigstrut}
\usepackage{braket}
\usepackage{color}
\usepackage{multirow}
\usepackage{mathtools}
\usepackage{dsfont}
\usepackage{float}
\usepackage[caption = false]{subfig}
\usepackage{colortbl}
\usepackage[colorlinks=true, citecolor=blue, urlcolor=blue]{hyperref}

\usetikzlibrary{arrows.meta,decorations.pathmorphing,positioning}
\usepackage{quantikz}
\newcommand{\nl}{\wireoverride{n}}

\definecolor{InputGrey}{HTML}{F2F3F5}
\definecolor{FourierOrange}{HTML}{F6D6A8}
\definecolor{BellBlue}{HTML}{DCECF7}
\definecolor{BellGateBlue}{HTML}{C5DFF1}
\definecolor{CorrectionGreen}{HTML}{D8EAD2}
\definecolor{OutputGreen}{HTML}{EEF7EA}
\definecolor{AlignmentOrange}{HTML}{F7E0C3}%
\definecolor{MeasurementBlue}{HTML}{DCEAF7}%
\definecolor{OutcomeGray}{HTML}{ECEFF1}%
\providecommand{\mathds}[1]{\mathbf{#1}}%
\tikzset{%
  system rail/.style={draw=black,line width=0.9pt},
  process arrow/.style={-{Latex[length=2.2mm,width=1.5mm]},draw=black,line width=0.9pt},
  classical arrow/.style={
  draw=black,
  double=white,
  double distance=1.2pt,
  line width=0.45pt,
  -{Latex[length=2.2mm,width=1.5mm]}
},
  entangled link/.style={draw=black,line width=0.9pt,decorate,decoration={snake,amplitude=0.65mm,segment length=2.4mm}},
  alignment gate/.style={draw=black,fill=AlignmentOrange,line width=0.9pt,minimum width=1cm,minimum height=0.9cm,inner sep=2pt},
  measurement/.style={draw=black,fill=MeasurementBlue,line width=0.9pt,minimum width=6cm,minimum height=1.5cm,align=center,inner sep=5pt},
  outcome/.style={draw=black,fill=OutcomeGray,line width=0.9pt,minimum width=0.8cm,minimum height=0.80cm,inner sep=2pt}
}

\newcommand{\be}{\begin{equation}}
\newcommand{\ee}{\end{equation}}
\newcommand{\ba}{\begin{eqnarray}}
\newcommand{\ea}{\end{eqnarray}}
\newcommand{\ketbra}[2]{|#1\rangle \langle #2|}

\newcommand{\tr}{\operatorname{Tr}}
\providecommand{\proj}[1]{\ket{#1}\!\bra{#1}}

\theoremstyle{definition}
\newtheorem{theorem}{Theorem}
\newtheorem{corollary}{Corollary}
\newtheorem{remark}{Remark}
\newtheorem{definition}{Definition}
\newtheorem{lemma}{Lemma}

\newcommand{\C}{\mathcal{C}}

\def\>{\rangle}
\def\<{\langle}

\usepackage[normalem]{ulem}
\usepackage{centernot}
\usepackage{subfig}
\usepackage{filecontents}

\providecommand{\ket}[1]{| #1{\rangle}}
\providecommand{\bra}[1]{\langle #1|}

\providecommand{\proj}[1]{\ket{#1}\bra{#1}}

\providecommand{\I}{\mathds{1}}
\providecommand{\dg}{^\dagger}
\providecommand{\nn}{\nonumber}
\providecommand{\abs}[1]{\left|#1\right|}

\newcommand{\sv}{\operatorname{sv}}

\providecommand{\C}{\mathbb{C}}

\providecommand{\zd}{\mathbb{Z}_d}
\providecommand{\zdx}{\mathbb{Z}_d^\times}
\providecommand{\op}[2]{|#1\rangle\!\langle#2|}

\begin{document}

\title{Entanglement-Swapping Measurements for Deterministic Entanglement Distribution}

\author{Mir Alimuddin}
\thanks{These authors contributed equally to the manuscript.}
\affiliation{ICFO - Institut de Ciencies Fotoniques, The Barcelona Institute of Science and Technology, 08860 Castelldefels, Barcelona, Spain}

\author{Jaemin Kim}
\email{jaemink@es.aau.dk}
\thanks{These authors contributed equally to the manuscript.}
\affiliation{Department of Electronic Systems, Aalborg University, Fredrik Bajers Vej 7C, 9220 Aalborg, Denmark}
\affiliation{School of Electrical Engineering, Korea Advanced Institute of Science and Technology (KAIST), 291 Daehak-ro, Yuseong-gu, Daejeon $34141$, Republic of Korea}

\author{Antonio Ac\'in}
\affiliation{ICFO - Institut de Ciencies Fotoniques, The Barcelona Institute of Science and Technology, 08860 Castelldefels, Barcelona, Spain}
\affiliation{ICREA - Instituci\'{o} Catalana de Recerca i Estudis Avan\c{c}ats, 08010 Barcelona, Spain}

\author{Leonardo Zambrano}
\thanks{These authors contributed equally to the manuscript.}
\affiliation{ICFO - Institut de Ciencies Fotoniques, The Barcelona Institute of Science and Technology, 08860 Castelldefels, Barcelona, Spain}

\begin{abstract}
Entanglement swapping is a key primitive for distributing entanglement over quantum networks, but different measurement outcomes can produce end-to-end states with different entanglement, requiring branch-dependent processing or the rejection of unfavorable outcomes.
We characterize all projective swapping measurements with full-Schmidt-rank vectors such that, for every pair of pure input links, all outcomes yield the same end-to-end state up to local-unitary corrections.
Within this family, the measurements that maximize the average G-concurrence for every input pair are built from complex Hadamard operators, and every outcome individually attains the optimum.
Classifying the underlying complex Hadamard operators that preserve optimal deterministic swapping gives one class for $d=2,3$, exactly $72$ classes for $d=5$, and uncountably many whenever $d=4k$. We show further that for $d=2,3$, the corrected end-to-end state in a swapping chain is independent of the swapping order, and discuss noise robustness under depolarizing noise and arbitrary convex input contamination.
For pure inputs, these schemes retain every outcome while achieving optimal G-concurrence and therefore eliminate outcome-based postselection.
\end{abstract}

\maketitle	

\section{Introduction}

Entanglement is a central resource of the second quantum revolution~\cite{Dowling'2003, Horodecki2009, Wehner'2018}. It enables tasks that have no classical counterpart or whose performance provably exceeds that of the best classical strategies, including secure key distribution~\cite{Ekert1991, Bennett1996, LoChau1999, ShorPreskill2000, Acin2007}, certified randomness generation~\cite{Pironio2010, ColbeckRenner2012, Ma2016,Herrero2017}, quantum computation~\cite{Raussendorf2001, JozsaLinden2003, Vidal2003, Raussendorf2003, Aaronson2004, Van2006, Gross2009, Bravyi2018}, quantum metrology~\cite{Giovannetti2004, Giovannetti2006, PezzeSmerzi2009, Demkowicz2015}, and enhanced classical and quantum communication~\cite{BennettWiesner1992, Bennett1993, BennettShor1999, BrunDevetak2006, Cubitt2010}. A central challenge is to distribute entanglement over long distances, where transmission loss and environmental noise degrade quantum correlations~\cite{Briegel1998,Pirandola2017}. Quantum networks address this challenge by establishing entangled links between neighboring nodes and connecting them through entanglement swapping at intermediate nodes~\cite{Zukowski1993, Dur1999, Pan1998, Duan2001, HardySong2000, AcinCiracLewenstein2007, CuquetCalsamiglia2009, PerseguersEtAl2010, PerseguersEtAl2013, MylavarapuEtAl2025, MengGaoHavlin2021, Sangouard2011, Azuma2015, GhosalEtAl2025}. In each swap, an intermediate node jointly measures one subsystem from each of two adjacent links. For generic inputs and measurements, the possible outcomes occur with different probabilities and leave the end parties with states having different Schmidt spectra. A protocol designed to produce a prescribed target state may therefore require branch-dependent entanglement processing or postselection on favorable outcomes~\cite{HardySong2000, AcinCiracLewenstein2007, CuquetCalsamiglia2009,PerseguersEtAl2010,PerseguersEtAl2013}. Although local filtering can increase the entanglement of the retained subensemble, an entanglement monotone cannot increase on average under local operations and classical communication (LOCC)~\cite{Horodecki1998,BrandaoPlenio2008,BrandaoPlenio2010,LamiRegula2023,HayashiYamasaki2025}. Discarding unfavorable outcomes therefore trades success probability for output quality and wastes the links consumed in the rejected branches.

This motivates the study of outcome-independent entanglement swapping: every measurement outcome should produce the same entangled state, up to local unitaries (LU) that can be corrected conditioned on the specific outcome. We consider two adjacent links sharing arbitrary pure states of local dimension $d$. After local unitary rotations align the incoming states in a common Schmidt basis, the intermediate node performs a complete rank-one projective measurement whose basis vectors have full Schmidt rank, as shown in Fig.~\ref{fig0}. We ask which measurements produce mutually LU-equivalent outputs for every pure input pair, a property we call \emph{universal LU-determinism}. Given the practical importance of resource efficiency, we also analyze optimality: under what additional conditions on the measurement is the entanglement of the final state optimal (in terms of some entanglement measure) for the given inputs?

Previous work has studied deterministic entanglement distribution or optimized pure-state networks under more specific objectives. For instance, Gour and Sanders constructed tailored measurements for which local corrections yield a single, outcome-independent state~\cite{Gour_2004}. In qubit networks, the $XZ$ Bell measurement basis has been used to maximize worst-case entanglement, yielding equiprobable outcomes with the same Schmidt spectrum~\cite{PerseguersEtAl2008}, and to achieve deterministic concurrence percolation~\cite{MengGaoHavlin2021}. 
In arbitrary local dimension, Meng et al. constructed a Fourier-based deterministic swapping rule for pure-state inputs, proved that it maximizes the average G-concurrence for a simple series network, and applied it to series-parallel quantum networks~\cite{MengEtAl2023}. 
Other studies have optimized ensemble-averaged quantities, including the average maximally entangled state yield along swapping chains~\cite{HardySong2000} and the propagation of average concurrence through sequential Bell measurements~\cite{BergouEtAl2021}. These results provide important constructions and optimality rules, but not a necessary-and-sufficient characterization of all projective measurements that are universally LU-deterministic.

In this work, we provide this characterization. We assume that nodes first apply local unitary rotations to align the incoming states to a common Schmidt basis. Under this alignment, we show that a full-Schmidt-rank projective measurement $\{|\Gamma_i \rangle \}$ is universally LU-deterministic if and only if its coefficient matrices $\{E_i\}$, defined by $|\Gamma_i \rangle = (E_i^* \otimes \mathds{1})|\Phi\rangle$, have uniform-modulus entries and belong to a common \emph{phase--conjugation} (PC) class: they are related by diagonal phase transformations ($E_j = D_L E_i D_R$), possibly combined with complex conjugation ($E_j = D_L E_i^* D_R$).

We then optimize the average output G-concurrence~\cite{Gour2004ConcurrenceMonotones}. This quantity is maximal for every pure input pair if and only if all measurement vectors $|\Gamma_i\rangle$ are maximally entangled. Combining this optimality condition with universal LU-determinism restricts the coefficient matrices to scaled complex Hadamard unitaries belonging to a common PC class. For these measurements, each conditional output has the optimal G-concurrence, and optimality is not merely an ensemble-average property.

This characterization leads to a dimension-dependent classification of the individual complex Hadamard coefficient matrices. There is a single PC class for $d=2,3$, exactly $72$ classes for $d=5$, and uncountably many whenever $d=4k$, where $k$ is a positive integer. The classification remains open for dimensions $d\geq6$ that are not divisible by $4$~\cite{Tadej, szollosi2011construction}. 
Whenever two classes produce distinct Schmidt spectra for full-Schmidt-rank outputs from the same inputs, the corresponding outputs have the same optimal G-concurrence but are incomparable under deterministic LOCC~\cite{Nielsen1999}.

Finally, we examine robustness and composition along network paths. Exact LU-determinism survives independent bipartite depolarizing noise when the measurement operators are related by diagonal unitaries without complex conjugation. For a general joint input of the form $\rho=(1- \epsilon) \rho_{\mathrm{ideal}} + \epsilon\sigma_{\mathrm{noise}}$, exact LU-equivalence need not persist, but the outcome-averaged trace distance from the ideal conditional outputs is at most $\epsilon$. We also consider chains of swapping nodes implementing optimal universally LU-deterministic measurements. For $d=2,3$, the corrected end-to-end state is independent of the swapping order, whereas a counterexample for $d=4$ shows that order independence is not guaranteed beyond $d=3$. Together, these results establish postselection-free, G-concurrence-optimal entanglement-swapping schemes for pure-state quantum networks, together with precise guarantees on their robustness and composition.

\section{Framework}
We begin with an elementary quantum network consisting of three parties: Alice ($\mathsf{A}$), a swapping node ($\mathsf{N}$), and Bob ($\mathsf{B}$).
The swapping node holds two subsystems, $\mathsf{N_A}$ and $\mathsf{N_B}$, associated with its links to Alice and Bob, respectively.
Alice and the node share a bipartite pure state $\ket{\tilde{\psi}}_{\mathsf{A N_A}}\in\mathcal{H}_{\mathsf{A}}\otimes\mathcal{H}_{\mathsf{N_A}}$, while the node and Bob share another bipartite pure state $\ket{\tilde{\phi}}_{\mathsf{N_B B}}\in\mathcal{H}_{\mathsf{N_B}}\otimes\mathcal{H}_{\mathsf{B}}$.
All four local Hilbert spaces have dimension $d$.
The parties may apply local unitaries, and the node performs a rank-one projective measurement $M_{\mathsf{N}}=\{\ket{\Gamma_i}\bra{\Gamma_i}\}_{i=1}^{d^2}$ on $\mathcal{H}_{\mathsf{N_A}}\otimes\mathcal{H}_{\mathsf{N_B}}$.
Conditioned on outcome $i$, Alice and Bob share a pure output state $\ket{\eta_i}_{\mathsf{AB}}$.
The overall procedure is illustrated in Fig.~\ref{fig0}.

Our main objective is to identify the class of measurements
\(M_{\mathsf{N}}\) satisfying the following condition:
\begin{definition}[Universal LU-determinism] \label{cond:1}
The protocol with a projective measurement $M_{\mathsf{N}}$ is \emph{universally LU-deterministic} if, for every pure input pair $(\ket{\tilde{\psi}},\ket{\tilde{\phi}})$, all normalized output states $\ket{\eta_i}$ associated with outcomes of nonzero probability are LU-equivalent.
\end{definition}

For a fixed input pair, this condition allows outcome-dependent local unitaries on $\mathsf{A}$ and $\mathsf{B}$ to map every conditional output to a common representative.
The qualifier \emph{universal} means that the property holds for every pure input pair; it does not mean that the output is independent of the inputs.

\begin{figure}
\centering
\resizebox{\linewidth}{!}{
\begin{tikzpicture}[x=1cm,y=1cm,font=\large]
  \coordinate (A-top) at (0.55,8.85);
  \coordinate (NA-top) at (3.45,8.85);
  \coordinate (NB-top) at (7.35,8.85);
  \coordinate (B-top) at (10.25,8.85);

  \node[font=\large] at (A-top) {$\mathsf{A}$};
  \node[font=\large] at (NA-top) {$\mathsf{N_A}$};
  \node[font=\large] at (NB-top) {$\mathsf{N_B}$};
  \node[font=\large] at (B-top) {$\mathsf{B}$};

  \draw[system rail] (0.55,8.35) -- (0.55,-0.35);
  \draw[system rail] (3.45,8.35) -- (3.45,3.35);
  \draw[system rail] (7.35,8.35) -- (7.35,3.35);
  \draw[system rail] (10.25,8.35) -- (10.25,-0.35);

  \draw[entangled link] (0.55,7.62) -- (3.45,7.62);
  \draw[entangled link] (7.35,7.62) -- (10.25,7.62);
  \node[fill=white,inner sep=1.5pt] at (2.00,7.98) {$\ket{\tilde{\psi}}$};
  \node[fill=white,inner sep=1.5pt] at (8.80,7.98) {$\ket{\tilde{\phi}}$};

  \node[alignment gate] (UA) at (0.55,6.38) {$U_1^\dagger$};
  \node[alignment gate] (VNA) at (3.45,6.38) {$V_1^*$};
  \node[alignment gate] (VNB) at (7.35,6.38) {$V_2^\dagger$};
  \node[alignment gate] (UB) at (10.25,6.38) {$U_2^*$};

  \draw[entangled link] (0.55,5.05) -- (3.45,5.05);
  \draw[entangled link] (7.35,5.05) -- (10.25,5.05);
  \node[fill=white,inner sep=1.5pt] at (2.00,5.42) {$\ket{\psi}$};
  \node[fill=white,inner sep=1.5pt] at (8.80,5.42) {$\ket{\phi}$};
  \node[text=black!65,font=\normalsize] at (5.40,5.05) {Schmidt-aligned inputs};

  \node[measurement] (measurement) at (5.40,3.55) {%
    \textbf{Swapping measurement}\\[2pt]%
    $\displaystyle M_{\mathsf N}=\bigl\{\ket{\Gamma_i}\!\bra{\Gamma_i}\bigr\}_{i=1}^{d^2}$%
  };
  \draw[process arrow] (3.45,4.75) -- (3.45,4.30);
  \draw[process arrow] (7.35,4.75) -- (7.35,4.30);

  \node[outcome] (outcome) at (5.40,1.95) {$i$};
  \draw[process arrow] (measurement.south) -- (outcome.north);
  \draw[classical arrow] (outcome.west) -- (0.55,1.95);
  \draw[classical arrow] (outcome.east) -- (10.25,1.95);
  \node[fill=white,text=black!65,inner sep=1.5pt,font=\normalsize] at (5.40,1.27) {outcome communication};

  \draw[entangled link] (0.55,0.42) -- (10.25,0.42);
  \node[fill=white,inner sep=1.5pt] at (5.40,0.03) {$\ket{\eta_i}$};
  \draw[process arrow] (0.55,0.85) -- (0.55,-0.35);
  \draw[process arrow] (10.25,0.85) -- (10.25,-0.35);
\end{tikzpicture}%
}
\caption{\textit{Entanglement-swapping protocol.} Alice and Bob initially share arbitrary pure states $\ket{\tilde{\psi}}_{\mathsf{A N_A}}$ and $\ket{\tilde{\phi}}_{\mathsf{N_B B}}$ with the swapping node. Local unitaries transform them into the Schmidt-aligned states $\ket{\psi}_{\mathsf{A N_A}}$ and $\ket{\phi}_{\mathsf{N_B B}}$, with their Schmidt coefficients reordered in the computational basis. The node then performs the projective measurement $M_{\mathsf N}=\{\ket{\Gamma_i}\!\bra{\Gamma_i}\}_{i=1}^{d^2}$. Conditioned on outcome $i$, communicated to the end parties, Alice and Bob share $\ket{\eta_i}_{\mathsf{AB}}=(AE_iB\otimes\mathds{1})\ket{\Phi}_{\mathsf{AB}}/\sqrt{p_i}$ with probability $p_i$.}
\label{fig0}
\end{figure}

To analyze this setting, we use the coefficient-matrix representation of bipartite vectors.
Any possibly unnormalized bipartite vector $\ket{\chi}$ can be represented by a complex matrix $X$ through $\ket{\chi}=(X\otimes\I)\ket{\Phi}$, where $\ket{\Phi}=\sum_{k=0}^{d-1}\ket{kk}$ is an unnormalized maximally entangled state~\cite{watrous2018theory}.
We call $X$ the \textit{coefficient matrix} of $\ket{\chi}$.
In this notation,
\begin{align}
\ket{\tilde{\psi}}
&=(\tilde{A}\otimes\I)\ket{\Phi}_{\mathsf{A N_A}},
\label{eq:input_state_A}\\
\ket{\tilde{\phi}}
&=(\tilde{B}\otimes\I)\ket{\Phi}_{\mathsf{N_B B}},
\label{eq:input_state_B}\\
\ket{\Gamma_i}
&=(E_i^*\otimes\mathds{1})\ket{\Phi}_{\mathsf{N_A N_B}}.
\label{eq:measurement}
\end{align}
The input normalization gives 
\begin{align}
\braket{\tilde{\psi}|\tilde{\psi}}
=\tr(\tilde{A}^\dagger\tilde{A})
=\braket{\tilde{\phi}|\tilde{\phi}}
=\tr(\tilde{B}^\dagger\tilde{B})
=1.
\end{align}
We write the measurement coefficients as $E_i^*$, the complex conjugates of the operators $E_i$, for later convenience.
The measurement vectors form an orthonormal basis, which in the coefficient-matrix representation gives
\begin{equation}
\tr(E_i^T E_j^*)=\delta_{ij}.
\end{equation}
For the projective measurement $M_{\mathsf{N}}=\{\proj{\Gamma_i}\}_{i=1}^{d^2}$, we refer to $\mathcal{O}(M_{\mathsf{N}})=\{E_i^*\}_{i=1}^{d^2}$ as the \textit{operator basis of the measurement} $M_{\mathsf{N}}$.
In the following characterization, we restrict to projective measurements whose vectors all have full Schmidt rank, or equivalently, whose associated matrices $E_i$ are nonsingular.

We next use the parties' local-unitary freedom to align both inputs in the computational Schmidt basis.
Write the singular-value decompositions as $\tilde{A}=U_1AV_1$ and $\tilde{B}=V_2BU_2$, where $A$ and $B$ are nonnegative real diagonal matrices.
As shown in Appendix~\ref{app:protocol}, Alice and Bob apply $U_1^\dagger$ and $U_2^*$ to $\mathcal{H}_{\mathsf{A}}$ and $\mathcal{H}_{\mathsf{B}}$, respectively, while the node applies $V_1^*\otimes V_2^\dagger$ to $\mathcal{H}_{\mathsf{N_A}}\otimes\mathcal{H}_{\mathsf{N_B}}$.
These local unitaries preserve the input entanglement, so the analysis reduces without loss of generality to
\begin{align}
\label{inputstates}
\ket{\psi}
&=(A\otimes\I)\ket{\Phi}_{\mathsf{A N_A}},
\qquad
\ket{\phi}
=(B\otimes\I)\ket{\Phi}_{\mathsf{N_B B}},
\end{align}
where $\tr(A^2)=\tr(B^2)=1$.

When the node obtains outcome $i$, the unnormalized state on $\mathsf{AB}$ and the outcome probability are
\begin{equation}
\label{eq:output_eta_i}
\ket{\hat{\eta}_i}
=(AE_iB\otimes\I_{\mathsf{B}})\ket{\Phi}_{\mathsf{AB}},
\quad
p_i=\tr\!\left[B^2E_i^\dagger A^2E_i\right].
\end{equation}
For $p_i>0$, the corresponding normalized state is $\ket{\eta_i}=\ket{\hat{\eta}_i}/\sqrt{p_i}$.

\section{Main results}
\subsection{Universal LU-determinism}
We now examine how universal LU-determinism constrains the set of measurements $ M_\mathsf{N} $.

\begin{lemma}\label{a+b}
Let $M_\mathsf{N}$ be a projective measurement whose vectors all have full Schmidt rank.
If the protocol is universally LU-deterministic, then, for every pure input pair, the measurement outcome probabilities are uniform, i.e., $p_i=1/d^2$.
\end{lemma}

See Appendix \ref{app:same_amplitude} for proof.
These constraints forbid any information leakage about the states through the output statistics, irrespective of any dependence of the measurement operators on the states. This situation is reminiscent of quantum teleportation \cite{Bennett1993}, where no information leaks during the process. 
Lemma~\ref{a+b}, together with $\mathrm{tr}(E_i^T E_i^\ast)=1$, imposes a structural constraint: each operator element belongs to the set of \emph{unbiased measurement operators}
\begin{equation}
\mathcal{S}_{\mathrm{UBM}} := \left\{ E^\ast \in\mathbb{C}^{d\times d} \middle| \abs{\langle m|E^\ast|n\rangle}=\frac{1}{d} \text{ for all }m,n \right\}.
\label{eq:UBM}
\end{equation}
However, the unbiasedness condition is basis-dependent: if the input states as defined in Eq.~\eqref{inputstates} have Schmidt bases different from the computational basis in $\mathcal{H}_\mathsf{N_A}$ and $\mathcal{H}_\mathsf{N_B}$, then the measurement operators will, in general, not be unbiased in that basis. Consequently, the outcome states produced by such a fixed measurement will not be equivalent for all possible inputs. This shows that universal determinism necessarily requires local rotations at the nodes prior to the swapping measurements. 

Choosing arbitrary $E_i, E_j \in \mathcal{S}_\mathrm{UBM}$ does not guarantee that the resulting states $\ket{\eta_i}$ and $\ket{\eta_j}$ are LU-equivalent for an arbitrary input pair, since some freedom remains in their complex phases. That is, Lemma~\ref{a+b} provides a necessary but not sufficient condition for universal LU-determinism. We now state the necessary and sufficient conditions under which two elements in $\mathcal{S}_\mathrm{UBM}$ generate LU-equivalent output states universally.
\begin{theorem}[Condition for Universal LU-Determinism]
\label{thm:phase-conjugation}
Let $E_i,E_j\in\mathcal{S}_{\mathrm{UBM}}$ be nonsingular.
The corresponding output states $\ket{\eta_i}$ and $\ket{\eta_j}$ are LU-equivalent for every pure input pair $\{\ket{\tilde{\psi}},\ket{\tilde{\phi}}\}$ if and only if either
\begin{align}\label{Main}
\text{(i)}~E_i = D_L E_j D_R \quad \text{or} \quad \text{(ii)}~E_i = D_L E_j^* D_R,
\end{align}
where $D_L$ and $D_R$ are diagonal unitaries in the computational basis.
\end{theorem}
The intuition behind the proof is as follows. For Schmidt-aligned inputs, the output associated with $E_i$ is determined by the matrix $AE_iB$, and two outputs are LU-equivalent precisely when these matrices have the same singular values. The diagonal matrices $A$ and $B$ contain the Schmidt coefficients of the two inputs and can be varied independently. Since the entries of $E_i$ and $E_j$ already have the same modulus, requiring the singular values of $AE_iB$ and $AE_jB$ to coincide for every choice of $A$ and $B$ strongly restricts their remaining phase patterns. The only possibilities are that the phases differ by independent row and column phases, or by such phases after complex conjugation, giving the two relations in Eq.~\eqref{Main}. Conversely, either relation leaves the output singular values unchanged for every input pair and therefore guarantees LU-equivalent outputs. The complete proof is given in Appendix~\ref{app:pc_equivalence}.
We group all operators $E_i \in \mathcal{S}_\mathrm{UBM}$ that are connected to a given $E_0 \in \mathcal{S}_\mathrm{UBM}$ via relation (i) or (ii) in Eq.~\eqref{Main} into the same \textit{phase-conjugation class} (PC-class). 
Operators from different PC-classes are distinguished by at least one pure input pair, although special input pairs may still produce LU-equivalent outputs.

To construct a measurement $M_\mathsf{N}$ satisfying universal LU-determinism, one must select an operator basis  $ \mathcal{O}(M_\mathsf{N}) = \{E_i^\ast\}_{i=1}^{d^2}$ from the same PC-class such that the corresponding projectors $\{\ket{\Gamma_i}\bra{\Gamma_i}\}_{i=1}^{d^2}$ in Eq.~\eqref{eq:measurement} form a valid projective measurement. 
One possible construction, inspired by Refs.~\cite{Gour_2004, Gour2005Gconcurrence}, applies to arbitrary $E^\ast \in \mathcal{S}_\mathrm{UBM}$ and yields the measurement
\begin{equation}
\mathcal{O}(M_\mathsf{N}) = \{ E_{m,m'}^\ast = D_{m'} E^\ast D_{m,m'} \}^{d-1}_{m,m'=0},  
\end{equation}
where
\begin{align}
D_{m'} &= \sum_k e^{2\pi i m' k/d} \proj{k}, \\
D_{m,m'} &= \sum_k e^{2\pi i (d m + m')k/d^2} \proj{k}.
\end{align}

Note that the Schmidt vector of the final state in Eq.~\eqref{eq:output_eta_i} is represented by $\sv(dAE_i B)$, where $\sv(\cdot)$ denotes the vector of non-increasing ordered singular values of the given matrix and $d$ is multiplied for normalization.
Universal LU-determinism is satisfied when $\sv(dAE_i B)$ is independent of the measurement outcome $i$ for arbitrary inputs $A$ and $B$.

\subsection{G-concurrence optimality}
For dimensions $d>2$, such Schmidt vectors are generically not ordered by majorization~\cite{Marshall}, so that deterministic LOCC convertibility does not induce a canonical total order on the outputs. 
We therefore compare measurements through a scalar entanglement monotone.
\begin{definition}[G-concurrence~\cite{Gour2004ConcurrenceMonotones}] \label{def:Gconc}
For a normalized pure state $\ket{\chi} = (M \otimes \I)\ket{\Phi}$ in dimension $d$, the G-concurrence is defined as $\C_d(\ket{\chi}) := d\,\abs{\det M}^{2/d}$.
\end{definition}
Its determinant form is particularly suited to entanglement swapping, since the output coefficient matrix $AE_iB$ yields an exact multiplicative factorization of the input and measurement contributions to the average output G-concurrence.
Equal to $d$ times the geometric mean of the Schmidt probabilities, $\C_d$ reduces to the concurrence for $d=2$ and takes values from $0$ (when $M$ is rank-deficient) to $1$ (for maximally entangled states). 
The average output G-concurrence is $\overline{\C_d} := \sum_{i=1}^{d^2} p_i\, \C_d(\ket{\eta_i})$.
\begin{definition}[G-concurrence optimality] \label{cond:2}
A measurement $M_\mathsf{N}$ is \emph{optimal} if, for every pure input pair $\{\ket{\tilde{\psi}},\ket{\tilde{\phi}}\}$, it maximizes the average output G-concurrence $\overline{\C_d}$ over all complete projective measurements at the node.
\end{definition}

\begin{lemma}[G-concurrence factorization]\label{G-factorization}
Let $\ket{\psi}$ and $\ket{\phi}$ be the input states with coefficient matrices $A$ and $B$, and let $M_\mathsf{N}$ be any complete projective measurement whose vectors $\ket{\Gamma_i}$ are associated with operators $E_i^\ast$ as in Eq.~\eqref{eq:measurement}.
Then
\begin{align}
\overline{\C_d} &= \C_d(\ket{\psi})\, \C_d(\ket{\phi})\, \frac{1}{d^2} \sum_{i=1}^{d^2} \C_d(\ket{\Gamma_i}) \nonumber\\ 
&\le \C_d(\ket{\psi})\, \C_d(\ket{\phi}), \label{eq:avg_factorization}
\end{align}
where, for inputs of full Schmidt rank, equality holds if and only if every $\ket{\Gamma_i}$ is maximally entangled.
If the protocol is additionally universally LU-deterministic, then for every outcome $i$,
\begin{align}
\C_d(\ket{\eta_i}) = \C_d(\ket{\psi})\,\C_d(\ket{\phi})\,\C_d(\ket{\Gamma_i}) . \nn
\end{align}
\end{lemma}
The proof is provided in Appendix~\ref{app:optimal_g-conc}; the product upper bound in Eq.~\eqref{eq:avg_factorization} was previously established in Ref.~\cite{Gour2004ConcurrenceMonotones}.
Since the input pairs in Definition~\ref{cond:2} include full-Schmidt-rank pairs, a measurement is optimal if and only if all of its vectors $\ket{\Gamma_i}$ are maximally entangled.
Under universal LU-determinism, all outcomes share the common value $\overline{\C_d}$, so an optimal universally LU-deterministic protocol attains $\C_d(\ket{\psi})\,\C_d(\ket{\phi})$ in every outcome. 
In this multiplicative sense, the G-concurrence quantifies the entanglement ``throughput'' of the swapping node.

We denote the operator representatives of maximally entangled vectors by
\begin{align}
\mathcal{S}_\mathrm{MEM}:=\left\{E^\ast=\tfrac{1}{\sqrt d}U \text{ for some unitary $U$} \right\},
\end{align}
and call $M_\mathsf{N}$ a \emph{maximally entangled measurement} if $E_i^\ast\in\mathcal{S}_\mathrm{MEM}$ for all $i$.
By Lemma~\ref{G-factorization}, G-concurrence optimality is equivalent to $M_\mathsf{N}$ being a maximally entangled measurement.

\subsection{Complex Hadamard Measurements}
Universal LU-determinism independently requires $E_i^\ast\in\mathcal{S}_\mathrm{UBM}$, so we define
\begin{align}
\mathcal{S}_\mathrm{CHM}:=\mathcal{S}_\mathrm{UBM}\cap\mathcal{S}_\mathrm{MEM}=\left\{E^\ast \!=\!\tfrac{1}{\sqrt d}H: H\in\mathbb{H}_d\right\},
\end{align}
where $\mathbb{H}_d$ is the set of $d\times d$ complex Hadamard unitaries, and call $M_\mathsf{N}$ a \emph{complex Hadamard measurement} if $E_i^\ast\in\mathcal{S}_\mathrm{CHM}$ for all $i$. Hence a measurement is optimal and universally LU-deterministic if and only if it is a complex Hadamard measurement whose operators belong to a common PC class.

In the qubit case $d=2$, this optimality is strict: for any given inputs of full Schmidt rank, the output produced by an optimal universally LU-deterministic measurement can be deterministically converted by LOCC into the output of any non-optimal universally LU-deterministic measurement.

\begin{figure}[t]
\centering
\resizebox{\linewidth}{!}{
\begin{quantikz}[]
& \lstick{\footnotesize $\mathsf{A}~$} \makeebit[angle=-30]{$\ket{\psi}$} \nl & & & \gate[style={fill=CorrectionGreen}]{Z^m} & \rstick[4]{$\ket{\eta_{00}}$} \\
\lstick[2,label style={transform canvas={xshift=-1ex}},braces={transform canvas={xshift=-1ex}}]{\small $\begin{matrix}
\text{Swapping} \\ \text{Node $\mathsf{N}$}
\end{matrix}$} & \lstick{\footnotesize $\mathsf{N_A}~$} \nl & \gate[style={fill=FourierOrange}]{F_d^\dagger} & \gate[2,style={fill=BellBlue}]{\begin{matrix} \text{Bell measurement} \\ \text{outcome $(m,n)$} \end{matrix}} & \ctrl[vertical wire=c]{-1} \setwiretype{c} \\
& \lstick{\footnotesize $\mathsf{N_B}~$} \makeebit[angle=-30]{$\ket{\phi}$} \nl & &   & \ctrl[vertical wire=c]{1} \setwiretype{c} \\
& \lstick{\footnotesize $\mathsf{B}~$} \nl & & & \gate[style={fill=CorrectionGreen}]{Z^n} &
\end{quantikz}
}
\caption{\textit{Circuit realization of a Fourier-class complex Hadamard measurement.} The input links are shown after local alignment in the computational Schmidt basis. The fixed pre-rotation $F_d^\dagger$ on $\mathsf{N_A}$ converts a conventional generalized Bell measurement into the Fourier-class measurement with $E_{mn}^{*} = \tfrac{1}{\sqrt d} Z^{m}F_dZ^{n}$. The Bell-measurement outcome $(m,n)$ is communicated to Alice and Bob, who apply $Z_{\mathsf A}^{m}$ and $Z_{\mathsf B}^{n}$, which map every conditional branch to $\ket{\eta_{00}}$; all outcomes occur with probability $1/d^2$ and are accepted.}
\label{fig:fourier-chm-circuit}
\end{figure}

The conventional Bell measurement is a maximally entangled measurement but is not a complex Hadamard measurement in the Schmidt-aligned basis. As illustrated in Fig.~\ref{fig:fourier-chm-circuit}, a single fixed local Fourier pre-rotation converts it into a Fourier-class complex Hadamard measurement, after which outcome-dependent local phase corrections map every branch to the same output state. This construction preserves G-concurrence optimality while adding universal LU-determinism in every dimension (see Appendix~\ref{app:bell-to-chm}).

\paragraph*{Qubit illustration.}
Consider the identical non-maximally entangled inputs
\begin{align}
\ket{\psi}_{\mathsf{A N_A}}=\ket{\phi}_{\mathsf{N_B B}}=\frac{\sqrt{3}}{2}\ket{00}+\frac{1}{2}\ket{11}, \nn\\
\C_2(\ket{\psi})=\C_2(\ket{\phi})=\frac{\sqrt{3}}{2}. \label{eq:example}
\end{align}
The conventional Bell measurement uses
\begin{align}
\ket{\phi^\pm}=\frac{\ket{00}\pm\ket{11}}{\sqrt{2}},
\qquad
\ket{\psi^\pm}=\frac{\ket{01}\pm\ket{10}}{\sqrt{2}},
\end{align}
and produces the normalized conditional outputs
\begin{align}
\ket{\eta_{\phi^\pm}}&=\frac{3\ket{00}\pm\ket{11}}{\sqrt{10}},
&
p_{\phi^\pm}&=\frac{5}{16},
&
\C_2(\ket{\eta_{\phi^\pm}})&=\frac{3}{5},
\\
\ket{\eta_{\psi^\pm}}&=\frac{\ket{01}\pm\ket{10}}{\sqrt{2}},
&
p_{\psi^\pm}&=\frac{3}{16},
&
\C_2(\ket{\eta_{\psi^\pm}})&=1.
\end{align}
The $\phi$- and $\psi$-type outputs are not LU-equivalent, although the Bell measurement remains average-concurrence optimal:
\begin{align}
\overline{\C_2}^{\mathrm{Bell}}
=
2\left(\frac{5}{16}\right)\frac{3}{5}
+
2\left(\frac{3}{16}\right)
=
\frac{3}{4}.
\end{align}

A representative qubit complex Hadamard measurement is the graph-state-basis measurement
\begin{align}
\ket{G_{mn}}&=(Z^m\otimes Z^n)\ket{G}, \nn\\
\ket{G}&=\mathrm{CZ}\ket{+}\ket{+} =\frac{\ket{00}+\ket{01}+\ket{10}-\ket{11}}{2}, \label{eq:graph_meas}
\end{align}
where $m,n\in\{0,1\}$. This basis satisfies $\mathcal{B}_{G}=(H\otimes\I)\mathcal{B}_{\mathrm{Bell}}$, up to outcome labels and global phases, and is therefore a locally Hadamard-rotated Bell basis. Its conditional outputs are
\begin{align}
\ket{\eta_{mn}^{G}} ={}& \frac{3}{4}\ket{00} +(-1)^n\frac{\sqrt{3}}{4}\ket{01} +(-1)^m\frac{\sqrt{3}}{4}\ket{10} \nn\\
&{} + (-1)^{m+n+1}\frac{1}{4}\ket{11}, \label{eq:graph_output} \\
p_{mn} ={}& \frac{1}{4}.
\end{align}
All four outcomes are LU-equivalent and attain
\begin{align}
\C_2(\ket{\eta_{mn}^{G}})=\overline{\C_2}^{G}=\frac{3}{4}=\C_2(\ket{\psi})\C_2(\ket{\phi}).
\end{align}
Thus, both measurements attain the same optimal average concurrence, while the graph-state complex Hadamard measurement distributes it uniformly among LU-equivalent outcomes. 
For two maximally entangled inputs, every output of either measurement is maximally entangled, so the measurement orientations yield the same output LU class.
This graph-state basis is the $XZ$ basis used in concurrence percolation, where it realizes the deterministic series rule $\C_2(\ket{\eta_{mn}^{G}})=\C_2(\ket{\psi})\C_2(\ket{\phi})$ for every outcome~\cite{PerseguersEtAl2008,MengGaoHavlin2021,MengEtAl2023}.

\subsection{PC-classes in complex Hadamard measurements}
We now count the PC-classes to which the individual coefficient matrices of a complex Hadamard measurement may belong, equivalently the PC-classes of complex Hadamard unitaries in $\mathbb H_d$.
\begin{theorem}[Phase-Conjugation Classes of Complex Hadamard Unitaries] \label{thm:PC_class_count}
For local dimension $d$, the complex Hadamard unitaries in $\mathbb H_d$, equivalently the scaled coefficient matrices in $\mathcal{S}_\mathrm{CHM}$, have the following numbers of distinct PC-classes:
\begin{center}
\begin{minipage}{0.72\linewidth}
\centering
\begin{tabular}{c|c|c|c|c|c}
\hline
$d$ & $2$ & $3$ & $4$ & $5$ & $4 k$ \\ \hline
Number of classes & $1$ & $1$ & $\infty$ & $72$ & $\infty$ \\ \hline
\end{tabular}
\end{minipage}\hfill
\begin{minipage}{0.25\linewidth}
\footnotesize
\raggedright
where $k\in\mathbb{Z}_{>0}$.
\end{minipage}
\end{center}
\end{theorem}

A detailed proof is provided in Appendix~\ref{app:pc_class_num}.
Theorem~\ref{thm:PC_class_count} classifies individual complex Hadamard coefficient operators, rather than complete orthonormal operator bases or projective measurements.
Representatives of distinct PC-classes produce different output Schmidt vectors for at least one input pair, although special inputs may still yield LU-equivalent outputs.
For any such full-Schmidt-rank distinguishing input pair, the outputs have equal optimal G-concurrence but are incomparable under deterministic LOCC~\cite{Nielsen1999}.

\section{Noise robustness}
While our main results concern pure inputs, exact LU-determinism survives independent bipartite depolarizing noise when the measurement operators satisfy $E_i=D_L^{(i)}E_1D_R^{(i)}$ for diagonal unitaries $D_L^{(i)}$ and $D_R^{(i)}$.
For a general joint input of the form $\rho=(1-\epsilon)\rho_{\mathrm{ideal}}+\epsilon\sigma_{\mathrm{noise}}$, exact LU-equivalence need not persist, but the outcome-averaged trace distance from the ideal outputs satisfies $\overline{D}\leq\epsilon$. The derivations and precise conditions are given in Appendix~\ref{app:noise_robustness}.

\paragraph*{Qubit illustration.}
Consider the inputs and graph-state complex Hadamard measurement introduced in Eqs.~\eqref{eq:example} and \eqref{eq:graph_meas}, with
\begin{align}
\rho_\psi=\rho_\phi=\operatorname{diag}\left(\frac{3}{4},\frac{1}{4}\right),
\end{align}
and apply the same depolarizing admixture $r$ to both links. The normalized conditional output is
\begin{align}
\rho_{mn}^{(r)}={}&(1-r)^2\proj{\eta_{mn}^{G}} \nn\\
&{} +r(1-r)\left(\rho_\psi\otimes\frac{\mathds{1}_2}{2} +\frac{\mathds{1}_2}{2}\otimes\rho_\phi\right) +r^2\frac{\mathds{1}_4}{4},
\end{align}
where $\ket{\eta_{mn}^{G}}$ is defined in Eq.~\eqref{eq:graph_output} and $p_{mn}^{(r)}=1/4$. Because the noise terms are diagonal,
\begin{align}
\rho_{mn}^{(r)}=(Z^m\otimes Z^n)\rho_{00}^{(r)}(Z^m\otimes Z^n)^\dagger,
\end{align}
so all noisy outputs remain LU-equivalent under the same outcome-dependent corrections as the ideal outputs. For $r=0.05$, the ideal component of the joint input has weight $(1-r)^2=0.9025$, corresponding to $\epsilon=1-(1-r)^2=0.0975$, while direct evaluation gives
\begin{align}
\overline{D}
=\frac{1}{2}\left\|\rho_{mn}^{(0.05)}-\proj{\eta_{mn}^{G}}\right\|_1
\approx0.0678
<0.0975=\epsilon.
\end{align}

\section{Entanglement distribution along network paths}
Bipartite measurements constitute the operational core of many practical quantum network architectures \cite{AcinCiracLewenstein2007,CuquetCalsamiglia2009,PerseguersEtAl2010,PerseguersEtAl2013,MylavarapuEtAl2025,MengGaoHavlin2021}. 
They extend entanglement across multiple nodes through local joint measurements on two subsystems at a time. Our \emph{building-block} swapping module can therefore be embedded along chosen paths in arbitrary network topologies, including linear chains, two-dimensional lattices, and more complex networks.

A key property of this construction is its \emph{universal LU-determinism}. For any chosen path and fixed swapping strategy, once each repeater implements a measurement satisfying this property with the intermediate corrections, all measurement-outcome branches yield LU-equivalent corrected end-to-end states, irrespective of the entanglement spectra of the elementary links.
Furthermore, when each repeater implements a universally LU-deterministic complex Hadamard measurement, each swap is G-concurrence optimal, with the optimal value attained for every measurement outcome.
Operationally, the protocol is \emph{postselection-free}: every outcome is accepted without retry, and outcome-conditioned local-unitary corrections map all branches to a common Schmidt-basis representative.

In a network scheme, a natural question is whether the order in which intermediate nodes perform their swapping operations affects the final entangled state shared by the end users.
Reference~\cite{MengEtAl2023} established associativity of this Fourier-based series rule for $d\in\{2,3\}$ and showed that associativity can fail for $d>3$.
We state the corresponding order-independence result for a network path and provide a self-contained proof and an explicit $d=4$ counterexample:
\begin{remark}[Order independence in low dimensions]
Consider a network path
$\mathsf{A}\!-\!\mathsf{N}_1\!-\!\cdots\!-\!\mathsf{N}_L\!-\!\mathsf{B}$
in which each swapping node implements a universally LU-deterministic complex Hadamard measurement.
For $d\in\{2,3\}$, the corrected end-to-end state shared by $\mathsf{A}$ and $\mathsf{B}$ is order-independent: its final LU-equivalence class does not depend on the swapping order. 
In higher dimensions, each swap remains G-concurrence optimal, but order independence is not guaranteed.
Appendix~\ref{app:general_network} provides detailed statements and proofs, and a counterexample for $d=4$.
\end{remark}

\section{Discussion \& Outlook} 

We completely characterized universally LU-deterministic full-Schmidt-rank projective measurements: their coefficient matrices must be unbiased and belong to a common PC-class, and these conditions are also sufficient.
Imposing G-concurrence optimality, equivalently the maximally entangled measurement condition for full-Schmidt-rank inputs, restricts them to complex Hadamard measurements whose operators share a PC-class.
The dimension-dependent count of complex Hadamard measurement operator PC-classes is complete for $2\leq d\leq5$: one for $d=2,3$, uncountably many for $d=4$, and $72$ for $d=5$.
Beyond these dimensions, we establish uncountably many PC-classes whenever $d=4k$, while the general count remains open.

Our results have several practical implications for quantum networks.
Because all outcomes within a given class produce LU-equivalent states, each round generates an outcome-independent amount of entanglement and requires no postselection.
In contrast, a generic protocol along a path containing $N$ swapping nodes may yield up to $d^{2N}$ branches with different amounts of entanglement.
For tasks requiring a fixed target state, such branches may call for distinct entanglement processing; our protocol avoids this branch-specific processing beyond local-unitary corrections.
A second advantage of our framework is that it guarantees optimal entanglement distribution with respect to the G-concurrence, not only on average but for every outcome individually.
Our universal swapping scheme is explicitly compatible with a natural experimental architecture in which fixed measurement devices are combined with outcome-dependent local corrections.
For $d=2,3$, order independence also removes the algebraic requirement to prescribe a swapping order along the path.
It merits further investigation whether this flexibility can relax synchronization requirements and reduce coherence-time demands in specific physical architectures.
Furthermore, non-conjugate measurements retain universal LU-determinism under input depolarizing noise; under arbitrary convex input contamination, the outcome-averaged trace-distance deviation is at most the contamination weight after each swapping operation.

Looking forward, several directions naturally emerge. Exploring multipartite and higher-dimensional swaps could generalize the protocol to networks beyond simple bipartite links, enabling scalable entanglement distribution in large, complex topologies. Finally, the connection to complex Hadamard matrices suggests potential links to other areas of quantum information and communication theory.

\begin{acknowledgments}
We thank Grzegorz Rajchel-Mieldzioć, Some Sankar Bhattacharya, Manik Banik, and Joonwoo Bae for useful discussions and valuable comments. 
M.A., A.A., and L.Z.\ were supported by 
the Government of Spain (Severo Ochoa CEX2019-000910-S, FUNQIP, and European Union NextGenerationEU PRTR-C17.I1), 
Fundaci\'o Cellex, 
Fundaci\'o Mir-Puig,
Generalitat de Catalunya (CERCA program), 
the European Union (QSNP, 101114043; QuantERA project Veriqtas), 
the ERC Advanced Grant CERQUTE, 
and the AXA Chair in Quantum Information Science.
M.A.\ additionally acknowledges funding from the European Union (QURES, 101153001).
J.K.\ acknowledges support from the Institute for Information \& Communications Technology Promotion (IITP) grants RS-2023-00229524,
RS-2025-02304540, and RS-2025-25464876, 
and from the Danish National Research Foundation (DNRF) through the Center of Excellence CLASSIQUE, grant no.\ DNRF187.
\end{acknowledgments}

\vspace{5pt}

\twocolumngrid
\bibliography{bib}

\clearpage
\onecolumngrid

\section*{Appendix}
\section{The protocol} \label{app:protocol}
A bipartite pure state $\ket{s}$ in a $d \times d$ Hilbert space and a $d \times d$ matrix $S$ have the following one-to-one correspondence:
\begin{align}
\ket{s} = \sum_{ij} s_{ij} \ket{ij} \leftrightarrow S = \sum_{ij} s_{ij} \ketbra{i}{j}
\end{align}
Note that $\proj{s}$ can be seen as a Choi matrix of $S$ as a filtering operator:
\begin{align}
\proj{s} &= (S \otimes \I) \proj{\Phi} (S\dg \otimes \I), \\
\ket{s} &= (S \otimes \I) \ket{\Phi},
\end{align}
where $\ket{\Phi} = \sum_{k=0}^{d-1} \ket{kk}$ is the unnormalized $d$-dimensional maximally entangled state.

The initial states of the protocol are
\begin{align}
\ket{\tilde{\psi}} &= (\tilde{A}\otimes\mathds{1})\ket{\Phi}_\mathsf{A N_A} = (U_1 A V_1 \otimes \mathds{1}) \ket{\Phi}_\mathsf{A N_A}, \\
\ket{\tilde{\phi}} &= (\tilde{B} \otimes \mathds{1}) \ket{\Phi}_\mathsf{N_B B} = (V_2 B U_2 \otimes \mathds{1}) \ket{\Phi}_\mathsf{N_B B},
\end{align}
where we have used the singular value decompositions of $\tilde{A}$ and $\tilde{B}$. The matrices $A$ and $B$ are real and diagonal, and the normalization $\operatorname{tr}(\tilde{A}^\dagger\tilde{A})=\operatorname{tr}(\tilde{B}^\dagger\tilde{B}) = 1$ implies $\Vert A \Vert_\mathrm{F} = \Vert B \Vert_\mathrm{F} = 1$.  

In the first step of the protocol, Alice, Bob, and the node apply the unitary $U_1\dg \otimes V_1^\ast \otimes V_2\dg \otimes U_2^\ast$ to the shared state $\ket{\tilde{\psi}} \otimes \ket{\tilde{\phi}}$ on the system $\mathsf{A N_A N_B B}$. The full state of the system then becomes
\begin{align}
(U_1\dg \otimes V_1^\ast \otimes V_2\dg \otimes U_2^\ast) \ket{\tilde{\psi}} \otimes \ket{\tilde{\phi}}
&= (U_1\dg \otimes V_1^\ast \otimes V_2\dg \otimes U_2^\ast)(U_1 A V_1 \otimes \I \otimes V_2 B U_2 \otimes \I)\ket{\Phi}_\mathsf{A N_A} \otimes \ket{\Phi}_\mathsf{N_B B} \nonumber \\
&= (A \otimes \I \otimes B \otimes \I) \ket{\Phi}_\mathsf{A N_A} \otimes \ket{\Phi}_\mathsf{N_B B} \nonumber \\
&= \ket{\psi} \otimes \ket{\phi}
\end{align}
where we have used the identity $\mathds{1} \otimes X \ket{\Phi} = X^T \otimes \mathds{1} \ket{\Phi}$. 

Thus, we can reduce the problem to the study of the protocol with initial states
\begin{align}
\ket{\psi} = (A \otimes \mathds{1})\ket{\Phi}, \qquad \ket{\phi} = (B \otimes \mathds{1})\ket{\Phi},
\end{align}
for diagonal and positive semidefinite $A$ and $B$ satisfying $\Vert A \Vert_\mathrm{F} = \Vert B \Vert_\mathrm{F} = 1$.

In the next step, the node performs a projective measurement $M_\mathsf{N} = \{M_i\}_{i=1}^{d^2}$ with elements
\begin{align}
M_i = \mathds{1}_\mathsf{A} \otimes \ket{\Gamma_i}\bra{\Gamma_i} _\mathsf{N_A N_B} \otimes \mathds{1}_\mathsf{B},
\quad\mathrm{where}~
\ket{\Gamma_i} = (E_i^*\otimes\mathds{1})\ket{\Phi}_\mathsf{N_A N_B},
\end{align}
normalized such that $\operatorname{tr}(E_i^T E_j^*)= \delta_{ij}$. After obtaining outcome $i$ at the node, the (unnormalized) state shared by Alice and Bob is
\begin{align}
\ket{\hat{\eta}_i}
&= \bra{\Gamma_i}_\mathsf{N_A N_B} \left(A \otimes \mathds{1} \otimes B \otimes \mathds{1} \right) \ket{\Phi}_\mathsf{A N_A} \otimes \ket{\Phi}_\mathsf{N_B B} \nonumber\\
&= \bra{\Phi}_\mathsf{N_A N_B} \left(A \otimes E_i^T \otimes B \otimes \mathds{1} \right) \ket{\Phi}_\mathsf{A N_A} \otimes \ket{\Phi}_\mathsf{N_B B} \nonumber\\
&= \bra{\Phi}_\mathsf{N_A N_B} ({A E_i B} \otimes \mathds{1} \otimes \mathds{1} \otimes \mathds{1}) \ket{\Phi}_\mathsf{A N_A} \otimes \ket{\Phi}_\mathsf{N_B B} \nonumber \\
&= ({A E_i B} \otimes \mathds{1})\ket{\Phi}_\mathsf{AB}, \label{eq:output_un}
\end{align}
with probability
\begin{align}
p_i=\braket{\hat{\eta}_i|\hat{\eta}_i} =
\operatorname{tr}\big(B^2 E_i^\dagger A^2 E_i \big).
\end{align}
We study the requirements on the operator basis of the measurement $\mathcal{O}(M_\mathsf{N}) = \{E_i^*\}_{i=1}^{d^2}$ such that the protocol is \emph{universally LU-deterministic}, meaning that for every pure input pair $\{\ket{\tilde{\psi}},\ket{\tilde{\phi}}\}$, the normalized output states $\ket{\eta_i}$ on $\mathsf{AB}$ conditioned on different measurement outcomes are mutually LU-equivalent.

\section{The components of $E_i$ have the same modulus $\tfrac{1}{d}$} \label{app:same_amplitude}

Assuming that $A$ and $B$ are full-rank operators, equality of the ranks of $A E_i B$ and $A E_j B$ is a necessary condition for these operators to have the same singular values. Since multiplication by full-rank operators preserves rank, this implies that $E_i$ and $E_j$ must have the same rank.
Having operators $E_i$ with deficient rank would mean that the resulting states $\ket{\eta_i} = \ket{\hat{\eta}_i} /\sqrt{p_i}$ would immediately have Schmidt rank smaller than $d$, even if the initial states $\ket{\psi}$ and $\ket{\phi}$ have Schmidt rank equal to $d$. Therefore, from now on, we consider only full-rank measurements $\{E_i^*\}_{i=1}^{d^2}$.

Here we show that if the normalized output states $\{\ket{\eta_i}\}_{i=1}^{d^2}$ are all LU-equivalent, then the probabilities $p_i$ are uniform and all $E_i^*$ have entries of equal modulus $\tfrac{1}{d}$.

\begin{theorem}
    If $\ket{\eta_i} = \tfrac{1}{\sqrt{p_i}} \ket{\hat{\eta}_i}$ is LU-equivalent to $\ket{\eta_j} = \tfrac{1}{\sqrt{p_j}} \ket{\hat{\eta}_j}$ for all $i, j$, and all $\{E_i^*\}_{i=1}^{d^2}$ are non-singular, then the probabilities $p_i$ satisfy
    \[
    p_i = \frac{\abs{\det E_i}^{2/d}}{\sum_j \abs{\det E_j}^{2/d}}.
    \]
\end{theorem}

\begin{proof}
    If two states are LU-equivalent, their singular values must be equal. If the singular values of $X$ are equal to those of $Y$, then $\abs{\det X } = \abs{\det Y }$. Using this, we have
    \begin{align}
        \abs{\det\!\left(\frac{1}{\sqrt{p_i}}A E_i B\right)}^{2/d} &= \abs{\det\!\left(\frac{1}{\sqrt{p_j}}A E_j B\right)}^{2/d}, \\
       \frac{1}{p_i} \abs{\det A}^{2/d} \abs{\det E_i}^{2/d} \abs{\det B}^{2/d}  &=  \frac{1}{p_j} \abs{\det A}^{2/d} \abs{\det E_j}^{2/d} \abs{\det B}^{2/d}.
    \end{align}
    Hence,
    \begin{align}
     \frac{1}{p_i} \abs{\det E_i}^{2/d} = \frac{1}{p_j}  \abs{\det E_j}^{2/d} = C,
    \end{align}
    for some constant $C$. From $\sum_i p_i = 1$, we obtain $C = \sum_i \abs{\det E_i}^{2/d}$. Since $\abs{\det (E_i)} \neq 0$, each probability is therefore
    \begin{align}
        p_i = \frac{\abs{\det E_i}^{2/d}}{\sum_j \abs{\det E_j}^{2/d}}.
    \end{align}
    \end{proof}

This shows that the probabilities must be independent of the initial states if the resulting states are to be unitarily equivalent and all $\{E^\ast_i\}_{i=1}^{d^2}$ are full rank. However, the set $\{E^\ast_i\}_{i=1}^{d^2}$ itself may still depend on the initial states, as it must be chosen so that the output states of the protocol are LU-equivalent.

\begin{theorem}
Let $\ket{\eta_i} = \frac{1}{\sqrt{p_i}} (A E_i B \otimes \mathds{1}) \ket{\Phi}$ be a normalized quantum state, where each $E_i$ is non-singular and satisfies $\|E_i\|_F = 1$, and $A$ and $B$ are real diagonal matrices such that $\Vert A \Vert_\mathrm{F} =  \Vert B \Vert_\mathrm{F} = 1$.  
If, for any choice of $A$ and $B$, all states in the set $\{\ket{\eta_i}\}_{i=1}^{d^2}$ are unitarily equivalent to one another, then necessarily $p_i = \tfrac{1}{d^2}$ and all matrix elements $\langle m|E_i^\ast|n \rangle$ satisfy $\abs{\langle m|E_i^\ast|n \rangle} = \tfrac{1}{d}$ for all $i, m, n$.
\end{theorem}

\begin{proof}
From the previous theorem, we have
    \begin{align}
        p_j \abs{\det E_i}^{2/d} = p_i  \abs{\det E_j}^{2/d}, \\
        \tr[B E_j^\dagger A^2 E_j B] \abs{\det E_i}^{2/d} = \tr[B E_i^\dagger A^2 E_i B]  \abs{\det E_j}^{2/d}, \\
        \sum_{mn} b_m^2 a_n^2  \left( | \langle m|E_j|n \rangle|^2 \abs{\det E_i}^{2/d}  -  | \langle m|E_i|n \rangle|^2 \abs{\det E_j}^{2/d} \right) = 0.
    \end{align}
    This implies that $|\langle m|E_j|n \rangle|^2 \abs{\det E_i}^{2/d} = |\langle m|E_i|n \rangle|^2 \abs{\det E_j}^{2/d}$. Since $\Vert E_j \Vert_\mathrm{F}^2 = \sum_{mn} |\langle m|E_j|n \rangle|^2 = 1$, it follows that $\abs{\det E_i} = \abs{\det E_j}$ for all $i, j$. This in turn implies that
    \[
    p_i = \frac{1}{d^2},
    \]
    since $p_i = \frac{\abs{\det E_i}^{2/d}}{\sum_j \abs{\det E_j}^{2/d}}$.
    We then have
    \begin{align}
       p_i =  \sum_{mn} b_m^2 a_n^2  | \langle m|E_i|n \rangle|^2 = \frac{1}{d^2}.
    \end{align}
    Because this equality must hold for all $A$ and $B$, and since $\sum_{mn} b_m^2 a_n^2 = 1$, it follows that for all $i, m, n$,
    \[
    \abs{\langle m|E_i|n \rangle} = \abs{\langle m|E_i^\ast|n \rangle} = \frac{1}{d}.
    \]
\end{proof}

\section{Characterization of LU-equivalent measurement operators}
\label{app:pc_equivalence}

We now prove Theorem~\ref{thm:phase-conjugation}. We first establish an elementary rigidity property of weighted sums of phases.

\begin{lemma}
\label{lem:weighted_phase_sums}
Let $\theta_1,\ldots,\theta_d$ and $\varphi_1,\ldots,\varphi_d$ be real numbers. If
\begin{equation}
\left|\sum_{k=1}^d t_k e^{i\theta_k}\right|=\left|\sum_{k=1}^d t_k e^{i\varphi_k}\right|
\qquad\text{for every }t_k\geq 0,
\label{eq:weighted_phase_sums_assumption}
\end{equation}
then there exist $s\in\{+1,-1\}$ and $\alpha\in\mathbb{R}$ such that
\begin{equation}
\varphi_k=s\theta_k+\alpha\pmod{2\pi}
\qquad(k=1,\ldots,d).
\label{eq:weighted_phase_sums_conclusion}
\end{equation}
\end{lemma}

\begin{proof}
Squaring Eq.~\eqref{eq:weighted_phase_sums_assumption} and comparing the coefficients of $t_k t_l$ for $k\neq l$ gives
\begin{equation}
\cos(\theta_k-\theta_l)=\cos(\varphi_k-\varphi_l)
\qquad(k,l=1,\ldots,d).
\end{equation}
Hence the two families of real unit vectors
\begin{equation}
x_k=(\cos\theta_k,\sin\theta_k),\qquad y_k=(\cos\varphi_k,\sin\varphi_k)
\end{equation}
have the same Gram matrix. Therefore there is an isometry between their spans that maps each $x_k$ to $y_k$, and this isometry can be extended to an orthogonal transformation of $\mathbb{R}^2$. Every orthogonal transformation of $\mathbb{R}^2$ is either a rotation or a reflection followed by a rotation, which gives Eq.~\eqref{eq:weighted_phase_sums_conclusion}. This argument also covers the degenerate case in which the vectors span only a line.
\end{proof}

\begin{proof}[Proof of Theorem~\ref{thm:phase-conjugation}]
Write $U=E_j$ and $V=E_i$. Since $U,V\in\mathcal{S}_{\mathrm{UBM}}$,
\begin{equation}
|u_{rk}|=|v_{rk}|=\frac{1}{d}
\qquad\text{for all }r,k.
\label{eq:UV_flat_entries}
\end{equation}
For real diagonal $A$ and $B$, define the unnormalized reduced operators
\begin{equation}
H_U(A,B)=AUB^2U^\dagger A,\qquad H_V(A,B)=AVB^2V^\dagger A.
\label{eq:HU_HV}
\end{equation}
Equation~\eqref{eq:UV_flat_entries}, together with $\tr A^2=\tr B^2=1$, gives
\begin{equation}
\tr H_U(A,B)=\tr H_V(A,B)=\frac{1}{d^2}.
\label{eq:HU_HV_trace}
\end{equation}
The normalized reduced states are therefore $d^2H_U(A,B)$ and $d^2H_V(A,B)$, so the corresponding pure output states are LU-equivalent if and only if $H_U(A,B)$ and $H_V(A,B)$ have the same spectrum.

Assume first that the output states are LU-equivalent for every pure input pair, equivalently that $H_U(A,B)$ and $H_V(A,B)$ have the same spectrum for every real diagonal $A,B$ satisfying $\tr A^2=\tr B^2=1$. Multiplying $U$ or $V$ independently on the left and right by diagonal unitaries preserves this spectral condition because these unitaries commute with $A$ and $B$. Since all entries of $U$ and $V$ are nonzero, we may therefore dephase them independently so that every entry in their first rows and first columns equals $1/d$. We write
\begin{equation}
u_{rk}=\frac{1}{d}e^{i\theta_{rk}},\qquad v_{rk}=\frac{1}{d}e^{i\varphi_{rk}},
\end{equation}
with
\begin{equation}
\theta_{1k}=\theta_{r1}=\varphi_{1k}=\varphi_{r1}=0
\qquad\text{for all }r,k.
\label{eq:dephased_phases}
\end{equation}

Set $x_r=a_r^2$ and $y_k=b_k^2$. Equality of the spectra implies equality of the second spectral moments, and direct expansion gives
\begin{align}
\tr H_U(A,B)^2
&=\frac{1}{d^4}\sum_{r,q}x_r x_q\left|\sum_k y_k e^{i(\theta_{rk}-\theta_{qk})}\right|^2,
\label{eq:second_moment_U}\\
\tr H_V(A,B)^2
&=\frac{1}{d^4}\sum_{r,q}x_r x_q\left|\sum_k y_k e^{i(\varphi_{rk}-\varphi_{qk})}\right|^2.
\label{eq:second_moment_V}
\end{align}
Although $x$ and $y$ satisfy $\sum_r x_r=\sum_k y_k=1$, both expressions are homogeneous in $x$ and $y$, so equality extends to arbitrary nonnegative $x$ and $y$. For fixed $y$, the difference of Eqs.~\eqref{eq:second_moment_U} and \eqref{eq:second_moment_V} is a quadratic polynomial in $x$ that vanishes on the nonnegative orthant; comparing its coefficients therefore gives, for every pair of rows $r,q$ and every $y_k\geq0$,
\begin{equation}
\left|\sum_k y_k e^{i(\theta_{rk}-\theta_{qk})}\right|=\left|\sum_k y_k e^{i(\varphi_{rk}-\varphi_{qk})}\right|.
\label{eq:row_pair_phase_sum}
\end{equation}
Lemma~\ref{lem:weighted_phase_sums} then implies
\begin{equation}
\varphi_{rk}-\varphi_{qk}=s_{rq}(\theta_{rk}-\theta_{qk})+\alpha_{rq}\pmod{2\pi}
\label{eq:row_pair_rigidity}
\end{equation}
for all $k$, where $s_{rq}\in\{+1,-1\}$ and $\alpha_{rq}\in\mathbb{R}$.

Taking $q=1$ and using Eq.~\eqref{eq:dephased_phases}, the case $k=1$ gives $\alpha_{r1}=0\pmod{2\pi}$, so each row has a sign $s_r\in\{+1,-1\}$ satisfying
\begin{equation}
\varphi_{rk}=s_r\theta_{rk}\pmod{2\pi}
\qquad\text{for every }k.
\label{eq:row_signs}
\end{equation}
It remains to show that one global sign can be chosen. Suppose two rows have opposite signs, say $s_r=+1$ and $s_q=-1$. Applying Eq.~\eqref{eq:row_pair_rigidity} to this pair and using the first column gives $\alpha_{rq}=0\pmod{2\pi}$ and hence
\begin{equation}
\theta_{rk}+\theta_{qk}=s_{rq}(\theta_{rk}-\theta_{qk})\pmod{2\pi}
\qquad\text{for every }k.
\end{equation}
If $s_{rq}=+1$, then $2\theta_{qk}=0\pmod{2\pi}$ for every $k$, whereas if $s_{rq}=-1$, then $2\theta_{rk}=0\pmod{2\pi}$ for every $k$. Thus whenever two row signs differ, at least one of those rows is real. Complex conjugation leaves a real row unchanged, so the sign assigned to such a row in Eq.~\eqref{eq:row_signs} can be reversed without changing the row. All non-real rows must therefore share one sign, and every real row may be assigned that same sign. Consequently, in the dephased gauge,
\begin{equation}
V=U\qquad\text{or}\qquad V=U^*.
\end{equation}
Undoing the independent dephasings gives
\begin{equation}
V=D_LUD_R\qquad\text{or}\qquad V=D_LU^*D_R
\end{equation}
for diagonal unitaries $D_L,D_R$.

Conversely, suppose $V=D_LUD_R$. Since $A$, $B$, $D_L$, and $D_R$ are diagonal,
\begin{equation}
H_V(A,B)=D_LH_U(A,B)D_L^\dagger.
\end{equation}
If instead $V=D_LU^*D_R$, then
\begin{equation}
H_V(A,B)=D_LH_U(A,B)^*D_L^\dagger.
\end{equation}
Because $H_U(A,B)$ is Hermitian, $H_U(A,B)^*=H_U(A,B)^T$ has the same spectrum as $H_U(A,B)$. Thus $H_U(A,B)$ and $H_V(A,B)$ have the same spectrum for every real diagonal $A,B$, so the corresponding normalized pure output states have the same Schmidt coefficients and are LU-equivalent.
\end{proof}

\section{Optimality in terms of G-concurrence} \label{app:optimal_g-conc}
Recall that the (unnormalized) output state shared by Alice and Bob after obtaining measurement outcome $i$ at the node is written as $\ket{\hat \eta_i} = (A E_i B \otimes \I) \ket{\Phi}_\mathsf{AB}$ with outcome probability $p_i = \braket{\hat{\eta}_i|\hat{\eta}_i}$ and the normalized state $\ket{\eta_i} = \tfrac{1}{\sqrt{p_i}} \ket{\hat \eta_i}$.
Define
\begin{align}
\hat G_i = A E_i B, ~~ G_i = \frac{1}{\sqrt{p_i}} \hat G_i = p_i^{-1/2} A E_i B,
\end{align}
Therefore,
\begin{align}
\ket{\hat \eta_i} = \sqrt{p_i} \ket{\eta_i} &= (\sqrt{p_i} G_i \otimes \I) \ket{\Phi} \nn\\
&= (\hat G_i \otimes \I) \ket{\Phi} \nn\\
&= (A E_i B \otimes \I) \ket{\Phi}. \label{output}
\end{align}
Note that $G_i$ is normalized, while $\hat G_i$ is not:
\begin{align}
\tr[G_i\dg G_i] = 1, ~~ \tr[\hat G_i\dg \hat G_i] = p_i.
\end{align}

Recall that the G-concurrence of a normalized pure state $\ket{\chi} = (M \otimes \I) \ket{\Phi}$ in dimension $d$ is $\C_d(\ket{\chi}) = d \abs{\det M}^{2/d}$.
In particular, $\C_d(\ket{\Gamma_i}) = d \abs{\det E_i}^{2/d}$ for the measurement vectors $\ket{\Gamma_i} = (E_i^\ast \otimes \I) \ket{\Phi}$, since $\abs{\det E_i^\ast} = \abs{\det E_i}$.

\begin{lemma}[Determinant Arithmetic Mean-Geometric Mean inequality] \label{lem:det_AMGM}
Let $\sv(M)=(\sqrt{\lambda_1}, \ldots, \sqrt{\lambda_d})$, where $\tr[M\dg M] = \sum_i \lambda_i = 1$. From the Arithmetic Mean-Geometric Mean inequality, we have
\begin{align}
\abs{\det M}^{2/d} = (\prod_i \lambda_i)^{1/d} \le \frac{1}{d} \sum_i \lambda_i = \frac{1}{d} \tr[M\dg M],
\end{align}
where the equality holds if and only if $\lambda_i = \tfrac{1}{d}$ for all $i$, i.e., $(M \otimes\I)\ket{\Phi} = (U \otimes \I) \ket{\phi^+_d}$ for the normalized maximally entangled state $\ket{\phi^+_d}$ and some unitary $U$~\cite{Gour2004ConcurrenceMonotones}.
\end{lemma}

We utilize Lemma~\ref{lem:det_AMGM} and the fact that $\det(p M) = p^d \det(M)$ to show that the average output G-concurrence factorizes through the inputs and the measurement vectors, and is bounded by the product of the input G-concurrences:
\begin{align}
\sum_i p_i \C_d(\ket{\eta_i}) &= \sum_i p_i d \abs{\det G_i}^{2/d} \nn\\
&= \sum_i d p_i \abs{p_i^{-d/2} \det \hat G_i}^{2/d} \nn\\
&= \sum_i d \abs{\det(A E_i B)}^{2/d} \nn\\
&= \sum_i d \abs{\det A}^{2/d} \abs{\det E_i}^{2/d} \abs{\det B}^{2/d} \nn\\
&= \abs{\det A}^{2/d} \abs{\det B}^{2/d} \sum_i d \abs{\det E_i}^{2/d} \nn\\
&\le \abs{\det A}^{2/d} \abs{\det B}^{2/d} \sum_i \tr[E_i\dg E_i] \nn\\
&= \abs{\det A}^{2/d} \abs{\det B}^{2/d} d^2 \nn\\
&= \C_d(\ket{\psi}) \C_d(\ket{\phi}). \label{eq:avg_bound}
\end{align}
The inequality in Eq.~\eqref{eq:avg_bound} follows from Lemma~\ref{lem:det_AMGM} applied to $E_i$ with $\tr[E_i\dg E_i] = 1$.
For rank-deficient inputs both sides of Eq.~\eqref{eq:avg_bound} vanish, and the average G-concurrence does not discriminate between measurements.
For inputs of full Schmidt rank, i.e., $\det A \det B \neq 0$, equality in Eq.~\eqref{eq:avg_bound} holds if and only if $\sv(E_i) = (\tfrac{1}{\sqrt{d}}, \ldots, \tfrac{1}{\sqrt{d}})$ for all $i$, i.e., every measurement vector $\ket{\Gamma_i}$ is maximally entangled and $M_\mathsf{N}$ is a maximally entangled measurement.
Up to this point, universal LU-determinism has not been used.
If, in addition, the protocol is universally LU-deterministic, then $p_i = 1/d^2$ by Lemma~\ref{a+b}, so that $\C_d(\ket{\eta_i}) = \C_d(\ket{\psi})\, \C_d(\ket{\phi})\, \C_d(\ket{\Gamma_i})$ for every outcome $i$, and all conditional outputs share this common value by LU-invariance of the G-concurrence.
Moreover, the operators of a universally LU-deterministic measurement are unbiased, $\abs{\langle m | E_i^\ast | n \rangle} = 1/d$ (Appendix~\ref{app:same_amplitude}), and a unitary $\sqrt{d}\, E_i$ whose entries all have modulus $1/\sqrt{d}$ is a complex Hadamard unitary.
Hence an optimal universally LU-deterministic measurement consists of operators proportional to complex Hadamard unitaries, and its common output value attains the bound: $\C_d(\ket{\eta_i}) = \C_d(\ket{\psi})\, \C_d(\ket{\phi})$.

\section{Number of Phase--conjugation classes in each dimension} \label{app:pc_class_num}
In this section, we investigate the number of phase--conjugation classes in the set of complex Hadamard unitaries $\mathbb{H}_d$ in each dimension $d$.

\subsubsection{Definitions and notation}
Entrywise complex conjugation is denoted by $(\cdot)^\ast$.
A \emph{diagonal unitary} is a diagonal matrix whose diagonal entries have magnitude $1$.
A \emph{permutation matrix} has exactly one $1$ in each row and column (zeros elsewhere); in particular, it is unitary.
We write $\op{j}{k}$ for the matrix with a $1$ in position $(j,k)$ and $0$ elsewhere.

For $d\in\mathbb{N}$, we identify row and column indices with the ring
\[
\zd=\{0,1,\dots,d-1\},
\]
and let
\[
\zdx=\{a\in\zd:\gcd(a,d)=1\}
\]
denote its multiplicative unit group. Let $S_d$ denote the set of $d\times d$ permutation matrices.

\begin{definition}[Fourier matrix]
Fix $d\in\mathbb{N}$ and $\omega=e^{2\pi i/d}$. The unitary Fourier matrix $F_d\in\mathbb{C}^{d\times d}$ is
\[
F_d = \sum_{j,k\in\zd} \frac{1}{\sqrt{d}}\;\omega^{\,jk} \op{j}{k}.
\]
\end{definition}

We begin by introducing two equivalence relations for classification within $\mathbb{H}_d$~\cite{Tadej}.
\begin{definition}[Phase--conjugation and phase--permutation equivalence]
Let $H_0,H_1 \in \mathbb{H}_d$ be complex Hadamard unitaries.
\begin{itemize}
\item \textit{Phase--conjugation equivalence (PC-equivalence).}
We say that $H_0$ and $H_1$ are PC-equivalent, written $H_0 \sim_{\mathrm{PC}} H_1$, if there exist
diagonal unitaries $D_L,D_R$ such that
\[
  H_0 = D_L H_1 D_R
  \quad\text{or}\quad
  H_0 = D_L H_1^\ast D_R .
\]
The corresponding PC-class of $H$ is
$
  [H]_{\mathrm{PC}} := \{ H' \in \mathbb{H}_d : H' \sim_{\mathrm{PC}} H \}.
$
\item \textit{Phase--permutation equivalence (PP-equivalence).}
We say that $H_0$ and $H_1$ are PP-equivalent, written $H_0 \sim_{\mathrm{PP}} H_1$, if there exist
diagonal unitaries $D_L,D_R$ and permutation matrices $P_L,P_R$ such that
\[
  H_0 = D_L P_L H_1 P_R D_R .
\]
The corresponding PP-class of $H$ is
$
  [H]_{\mathrm{PP}} := \{ H' \in \mathbb{H}_d : H' \sim_{\mathrm{PP}} H \}.
$
\end{itemize}
\end{definition}

\begin{definition}[Permutation orbit of $F_d$]
The permutation orbit of $F_d$ under left and right multiplication by permutation matrices is
\[
[F_d]_\mathrm{P}
:= \{P_L F_d P_R : P_L,P_R\in S_d\}.
\]
This is a subset of the phase--permutation class $[F_d]_{\mathrm{PP}}$.
Since multiplying on the left and right by diagonal unitaries does not change phase--conjugation classes, the number of phase--conjugation classes inside the full phase--permutation class $[F_d]_{\mathrm{PP}}$ coincides with the number of such classes induced on the permutation orbit $[F_d]_\mathrm{P}$.
\end{definition}

\begin{definition}[Right coset]
Let $G$ be a group and $H \subseteq G$ a subgroup. For $g\in G$, the \emph{right coset} of $H$ with representative $g$ is
\[
gH := \{g h : h\in H\}.
\]
Right cosets of $H$ in $G$ partition $G$, and two elements $g,g'\in G$ lie in the same right coset iff $g^{-1}g'\in H$.
\end{definition}

\begin{definition}[Symmetry subgroup $G_d$]
The direct product $S_d\times S_d$ acts on matrices by
\[
(P_L,P_R^{-1})\cdot H:=P_LHP_R.
\]
Because permutation matrices are real and conjugate diagonal unitaries to diagonal unitaries, this action preserves PC-equivalence and descends to PC-classes. Define
\[
G_d:=\left\{(Q_L,Q_R^{-1})\in S_d\times S_d: \exists \text{diagonal unitaries }D_L,D_R\text{ with }Q_LF_dQ_R\in\{D_LF_dD_R,D_LF_d^\ast D_R\}\right\}.
\]
Thus, $G_d$ is the stabilizer of $[F_d]_{\mathrm{PC}}$ under this action and is therefore a subgroup of $S_d\times S_d$.
\end{definition}

\subsubsection{Fourier gate permutation orbit}\label{app:Fourier_PP_orbit}
We first focus on the permutation orbit of the Fourier gate.
\paragraph{Permutation as affine map}
\begin{lemma}[Characterization of permutation symmetries of $F_d$]\label{lem:affine}
Let $Q_L,Q_R$ be permutation matrices (row/column permutations). Then
\[
Q_L F_d Q_R \in \{D_L F_d D_R,\; D_L F_d^\ast D_R\}
\quad\text{for some diagonal unitaries }D_L,D_R
\]
if and only if there exist $\alpha,\gamma\in\zdx$ and $\beta,\delta\in\zd$ with
\[
Q_L = \op{j}{\alpha j+\beta},\qquad
Q_R = \op{\gamma k+\delta}{k},
\]
and $\alpha\gamma\equiv +1\pmod d$ in the first case, or $\alpha\gamma\equiv -1\pmod d$ in the second.
\end{lemma}

\begin{proof}
\textbf{\textit{If} direction. }
If $\bra j Q_L=\bra{\alpha j+\beta}$ and $Q_R \ket k=\ket{\gamma k+\delta}$ with $\alpha,\gamma\in\zdx$, then
\begin{align} \label{eq:permuted_Fd}
\braket{j|Q_L F_d Q_R|k}
= \frac{1}{\sqrt d}\,\omega^{(\alpha j+\beta)(\gamma k+\delta)}
= \omega^{\alpha\delta j} \omega^{\beta\gamma k} \omega^{\beta\delta} \cdot \frac{1}{\sqrt d}  \omega^{(\alpha\gamma)jk},
\end{align}
where $\omega^{\alpha\delta j}$ corresponds to the row phase, $\omega^{\beta\gamma k}$ corresponds to the column phase, and $\omega^{\beta\delta}$ corresponds to the global phase.
Thus $Q_L F_d Q_R = D_L F_d D_R$ when $\alpha\gamma\equiv 1\pmod d$ and $Q_L F_d Q_R = D_L F_d^\ast D_R$ when $\alpha\gamma \equiv -1\pmod d$, with diagonal unitaries collecting the row/column/global phases. 

\textbf{\textit{Only if} direction. }
Let $\sigma,\tau:\zd\to\zd$ be the bijections induced by $Q_L,Q_R$.
Then
\[
Q_LF_dQ_R= \sum_{j,k\in\zd} \frac{1}{\sqrt{d}}\;\omega^{\,\sigma(j)\,\tau(k)} \op{j}{k}.
\]
Assume first that $Q_LF_dQ_R=D_L F_d D_R$ (the conjugate branch is identical with $jk$ replaced by $-jk$).
Entrywise there exist functions $r,c:\zd\to\zd$ and a constant $c_0\in\zd$ such that
\begin{equation}\label{eq:phase-factorization}
\sigma(j)\,\tau(k)\equiv jk + r(j)+c(k)+c_0 \pmod d\qquad(j,k\in\zd).
\end{equation}

For fixed $s, t \in\zd$, applying a transformation
\[
f(j,k) \mapsto f(j+s, k+t) - f(j, k+t) - f(j+s, k) + f(j,k).
\]
to both sides of Eq.~\eqref{eq:phase-factorization} yields
\begin{equation}\label{eq:mixed}
\big(\sigma(j+s)-\sigma(j)\big)\,\big(\tau(k+t)-\tau(k)\big) \equiv s t \pmod d.
\end{equation}
Define the increments $D_\sigma(j;u):=\sigma(j+u)-\sigma(j)$ and
$D_\tau(k;u):=\tau(k+u)-\tau(k)$. Then \eqref{eq:mixed} reads
\begin{equation}\label{eq:prod}
D_\sigma(j;s)\cdot D_\tau(k;t)\equiv st \pmod d
\qquad \forall j,k,s,t\in\zd.
\end{equation}
Taking $s = t = 1$ leads to
\[
D_\sigma(j;1) D_\tau(k;1) \equiv 1 \pmod d
\qquad \forall j,k \in\zd.
\]
For each $j$, $D_\sigma(j;1)$ is a unit and for each $k$, $D_\tau(k;1)$ is a unit as well. 
(Note: an element $a \in \zd$ is a \textit{unit} iff there exists some $b \in \zd$ such that $ab \equiv 1 \pmod d$.)

Furthermore, for any $j_1, j_2, k \in \zd$, 
\[
\begin{aligned}
(D_\sigma(j_1;1) - D_\sigma(j_2;1)) D_\tau(k;1) \equiv 0 \pmod d \\
D_\sigma(j_1;1) \equiv D_\sigma(j_2;1) \pmod d
\end{aligned}
\]
since $D_\tau(k;1)$ is a unit, thus invertible. Consequently, $D_\sigma(j;1)$ is constant in $j$ and similarly, $D_\tau(k;1)$ is constant in $k$:
\begin{align} \label{eq:constant_delta}
D_\sigma(j; 1) = \alpha ~ \forall j, \quad D_\tau(k; 1) = \gamma ~ \forall k, ~~
\text{such that } \alpha \gamma \equiv 1 \pmod d.
\end{align}

From Eq.~\eqref{eq:prod} and Eq.~\eqref{eq:constant_delta}, for arbitrary $u \in \zd$,
\begin{align}
D_\sigma(j; u) D_\tau(k; 1) \equiv u \cdot 1 \pmod d, & \\
D_\sigma(j; u) \gamma \equiv u \pmod d. &
\end{align}
As shown in Eq.~\eqref{eq:constant_delta}, $\alpha, \gamma \in \zdx$ are invertible with $\gamma^{-1} \equiv \alpha$, thus
\begin{align*}
D_\sigma(j; u) \equiv \gamma^{-1}u \equiv \alpha u := f(u).
\end{align*}
Similarly,
\begin{align*}
D_\tau(k; u) \equiv \gamma u := g(u).
\end{align*}

Recalling the definition of $D_\sigma(j; u)$ and $D_\tau(k; u)$, 
\begin{align*}
D_\sigma(j; u) &= \sigma(j+u) - \sigma(j) = \alpha u, \\
D_\tau(k; u) &= \tau(k+u) - \tau(k) = \gamma u,
\end{align*}
the permutations $\sigma$ and $\tau$ are arithmetic progressions modulo $d$, hence affine:
\begin{align*}
\sigma(j) &= \sigma(0) + j\alpha = \alpha j + \beta, \\
\tau(k) &= \tau(0) + k\gamma = \gamma k + \delta,
\end{align*}
for some $\beta,\delta\in\mathbb Z_d$.

This proves that $\alpha \gamma \equiv +1 \pmod d$ for the case $Q_L F_d Q_R = D_L F_d D_R$. If instead $Q_LF_dQ_R=D_LF_d^\ast D_R$, the same argument holds with $jk$ replaced by $-jk$ in \eqref{eq:phase-factorization}, giving $\alpha\gamma\equiv -1\pmod d$.
This completes the proof.
\end{proof}

\paragraph{Symmetry group $G_d$ and orbit counting}

\begin{lemma}[Size of $G_d$]\label{lem:Gd-size}
Let $\varphi$ be Euler's totient function $\varphi(d) = \abs{\zdx}$ and define
\[
\varepsilon(d)=\begin{cases}
1,& d=1,2,\\
2,& d\ge 3.
\end{cases}
\]
Then 
\begin{align} \label{eq:size_G_d}
|G_d|=\varepsilon(d)\,d^2\,\varphi(d).
\end{align}
\end{lemma}

\begin{proof}
By Lemma~\ref{lem:affine}, the elements of $G_d$ are the pairs $(Q_L,Q_R^{-1})$ for which $Q_L$ and $Q_R$ induce affine permutations
\[
j\mapsto\alpha j+\beta,\qquad k\mapsto\gamma k+\delta,
\]
with $\alpha,\gamma\in\zdx$, $\beta,\delta\in\zd$, and $\alpha\gamma\equiv\pm1\pmod d$. For $d\ge3$, the two signs are distinct, whereas they coincide for $d=1,2$. The translations $\beta,\delta$ contribute $d^2$ choices, $\alpha$ contributes $\varphi(d)$ choices, and $\gamma$ is fixed by $\gamma\equiv(\pm1)\alpha^{-1}\pmod d$. Hence
\[
|G_d|=\varepsilon(d)\,d^2\,\varphi(d).
\]
\end{proof}

\begin{theorem}[Phase--conjugation classes in the Fourier permutation orbit]\label{thm:main-count}
Let $[F_d]_\mathrm{P}=\{P_LF_dP_R:\ P_L,P_R\in S_d\}$. The number $N_d$ of PC-classes induced on $[F_d]_\mathrm{P}$ is
\[
N_d=\frac{(d!)^2}{\varepsilon(d)\,d^2\,\varphi(d)},
\qquad
\varepsilon(d)=
\begin{cases}
1,&d=1,2,\\
2,&d\ge3.
\end{cases}
\]
For every $H\in[F_d]_\mathrm{P}$,
\[
\left|[H]_{\mathrm{PC}}\cap[F_d]_\mathrm{P}\right|=\varepsilon(d)\,d^2.
\]
The same $N_d$ counts the PC-classes in the full phase--permutation class $[F_d]_{\mathrm{PP}}$.
\end{theorem}

\begin{proof}
For
\[
g=(P_L,P_R^{-1}),\qquad g'=(P'_L,{P'_R}^{-1}),
\]
the corresponding matrices are $g\cdot F_d=P_LF_dP_R$ and $g'\cdot F_d=P'_LF_dP'_R$. By Lemma~\ref{lem:affine}, they are PC-equivalent if and only if
\[
g^{-1}g'
=
\left(P_L^{-1}P'_L,P_R{P'_R}^{-1}\right)
=
\left(P_L^{-1}P'_L,(P'_RP_R^{-1})^{-1}\right)
\in G_d.
\]
Equivalently, $g'\in gG_d$. Thus, the PC-classes induced on $[F_d]_\mathrm{P}$ correspond to the right cosets of $G_d$ in $S_d\times S_d$, and
\[
N_d=[S_d\times S_d:G_d]
=\frac{(d!)^2}{|G_d|}
=\frac{(d!)^2}{\varepsilon(d)\,d^2\,\varphi(d)}.
\]

The exact stabilizer of $F_d$ is
\[
S:=\left\{(Q_L,Q_R^{-1})\in S_d\times S_d:\ Q_LF_dQ_R=F_d\right\}\subseteq G_d.
\]
By Lemma~\ref{lem:affine}, exact equality requires $\beta=\delta=0$ and $\alpha\gamma\equiv1\pmod d$. Each $\alpha\in\zdx$ determines one such pair, with $\gamma=\alpha^{-1}$, so
\[
|S|=\varphi(d).
\]
For $H=g\cdot F_d$, the set $[H]_{\mathrm{PC}}\cap[F_d]_\mathrm{P}$ is the image of $gG_d$ under the orbit map, whose fibers are the right cosets of $S$. Therefore,
\[
\left|[H]_{\mathrm{PC}}\cap[F_d]_\mathrm{P}\right|
=\frac{|G_d|}{|S|}
=\varepsilon(d)\,d^2.
\]

Finally, every matrix $D_LP_LF_dP_RD_R\in[F_d]_{\mathrm{PP}}$ is PC-equivalent to $P_LF_dP_R\in[F_d]_\mathrm{P}$. Diagonal factors therefore introduce no additional PC-classes, so the same $N_d$ counts the PC-classes in $[F_d]_{\mathrm{PP}}$.
\end{proof}

\begin{corollary}[Small dimensions] \label{cor:N_d_table}
For $d=2,3,4,5,6$ the number $N_d$ of PC-classes in the PP-orbit of $F_d$ is
\begin{center}
\begin{tabular}{|c|c|c|c|c|c|}
\hline
$d$   & 2 & 3 & 4 & 5  & 6    \\
\hline
$N_d$ & 1 & 1 & 9 & 72 & 3600 \\
\hline
\end{tabular}    
\end{center}
\end{corollary}

\subsubsection{Proof for unique PC-class in $d\in\{2,3\}$} \label{app:pc_class_d=2,3}

\begin{remark} \label{rem:single class}
For $d \in \{2,3,5\}$, there exists only a unique PP-class of complex Hadamard unitaries $\mathbb{H}_d$ \cite{szollosi2011construction}. Any $H_i \in \mathbb{H}_d$ in each dimension $d \in \{2,3,5\}$ is PP-equivalent to the $d$-dimensional Fourier unitary $F_d$.
\end{remark}
In other words, in dimensions $d \in \{2,3,5\}$, any complex Hadamard unitary $U$ can be expressed as
\begin{align}
U = D_1 P_1 F_d P_2 D_2
\end{align}
for some permutation matrix $P_i$ and diagonal phase unitary $D_i$, where $F_d$ is the Fourier-unitary.

Combining Remark~\ref{rem:single class}, Theorem~\ref{thm:main-count}, and Corollary~\ref{cor:N_d_table} implies that there exists a unique PC-class in $\mathbb{H}_d$ for $d \in \{2,3\}$.

Additionally, note that for $d \in \{2,3\}$, any permutation $P_i$ can be expressed as a combination of a Weyl shift and a reversal:
\begin{align}
P_i = X^{s_i} R^{r_i}
\end{align}
for $s_i \in \{0, \ldots, d-1\}$ and $r_i \in \{0,1\}$.

We provide an alternative proof that a unique PC-class exists in $d \in \{2,3\}$ by showing the following remark. 
\begin{remark} \label{rem:perm_rev_Fourier}
Any complex Hadamard matrix $H$ that can be expressed as $H = D_1 P_1 F_d P_2 D_2$, where $P_i$ are cyclic permutation matrices with reversal $P_i=X^{s_i} R^{t_i}$ for $s_i\in\{0,\ldots,d-1\}, t_i\in\{0,1\}$ where $X:\ket{k}\mapsto\ket{k+1}$ and $R:\ket{k}\mapsto\ket{-k}$ and $F_d$ is the complex Fourier matrix, belongs to the PC-orbit of $F_d$.
\end{remark}
\begin{proof}
Define a Weyl shift operator $X$, a Weyl phase operator $Z$, and a reversal operator $R$:
\begin{align}
X &= \sum_{k=0}^{d-1} \ket{k+1~(\rm{mod}~d)} \bra{k}, \label{eq:Weyl_shift} \\
Z &= \sum_{k=0}^{d-1} \omega^k \proj{k}, ~ \omega=e^{2\pi i/d}, \label{eq:Weyl_phase} \\
R &= \sum_{k=0}^{d-1} \ket{-k~(\rm{mod}~d)} \bra{k}.
\end{align}
Suppose that the permutations $P_i$ can be expressed as a combination of a Weyl shift and a reversal:
\begin{align}
P_i = X^{s_i} R^{r_i}
\end{align}
for $s_i \in \{0, \ldots, d-1\}$ and $r_i \in \{0,1\}$.

From the fact that $Z^a X^b = \omega^{ab} X^b Z^a$, $RZ^s = Z^{-s} R$, $F_d X^s = Z^s F_d$, $F_d^\ast X^s = Z^{-s} F_d^\ast$, and $F_d R = R F_d = F_d^\ast$,
\begin{align}
U &= D_1 P_1 F_d P_2 D_2 \nonumber \\
&= D_1 X^{s_1} R^{r_1} F_d X^{s_2} R^{r_2} D_2 \nonumber \\
&= D_1 F_d^{\circ} Z^{\pm s_1} X^{s_2} R^{r_2} D_2 \nonumber \\
&= D_1 F_d^{\circ} \omega^{\pm s_1 s_2} X^{s_2} Z^{\pm s_1} R^{r_2} D_2 \nonumber \\
&= D_1 Z^{\pm s_2} F_d^{\circ} R^{r_2} \omega^{\pm s_1 s_2} Z^{\pm s_1} D_2 \nonumber \\
&= D_L F_d^{\circ} D_R,
\end{align}
where $F_d^{\circ} \in \{F_d, F_d^\ast\}$, $D_L = D_1 Z^{\pm s_2}$, and $D_R = \omega^{\pm s_1 s_2} Z^{\pm s_1} D_2$. The signs $\pm$ in the exponents of the Weyl operators and either $F_d^\circ=F_d$ or $F_d^\circ=F_d^\ast$ depend on the $r_1$ and $r_2$.
This completes the proof.
\end{proof}

\subsubsection{Proof for $72$ PC-classes in $d=5$}\label{d=5}
Combining Remark~\ref{rem:single class}, Theorem~\ref{thm:main-count}, and Corollary~\ref{cor:N_d_table} implies that there are exactly $72$ distinct PC-classes in $\mathbb{H}_5$. 

\subsection{Proof for infinite PC-classes in $d=4$}\label{d=4}

\begin{lemma}
There exists an infinite number of equivalence classes of complex Hadamard matrices in $d=4$ under the phase-conjugation operation.
\end{lemma}

\begin{proof}
Let $\sim_{PP}$ denote equivalence under phase or permutation, and $\sim_{PPC}$ denote equivalence under phase, permutation, or complex conjugation. We first show that there exists an infinite number of equivalence classes under phase--permutation--conjugation.

We consider the continuous, one-parameter family of complex Hadamard matrices $U(\alpha)$, defined for $\alpha \in \mathbb{R}$:
\begin{equation} \label{eq:dim4_ub}
U(\alpha) = \frac{1}{2}\begin{bmatrix}
1 & 1 & 1 & 1 \\
1 & i e^{i\alpha} & -1 & -i e^{i\alpha} \\
1 & -1 & 1 & -1 \\
1 & -i e^{i\alpha} & -1 & i e^{i\alpha}
\end{bmatrix}
\end{equation}
Under $\sim_{PP}$, the inequivalent classes of this family are parametrized by the interval $\alpha \in [0, \pi)$. Specifically, $U(\alpha) \sim_{PP} U(\alpha + \pi)$ \cite{Tadej}.

To determine the classes under $\sim_{PPC}$, we examine the action of complex conjugation on $U(\alpha)$. Taking the entry-wise complex conjugate gives
\[
U(\alpha)^*=U(\pi-\alpha).
\]
Thus, under the extended equivalence $\sim_{PPC}$, we have the identification
\[
U(\alpha) \sim_{PPC} U(\pi - \alpha)
\]
This identification maps the parameter space $[0, \pi)$ onto the reduced interval $[0, \pi/2)$. Since the interval $[0, \pi/2)$ is a continuum, we have an infinite number of inequivalent classes under phase--permutation-conjugation. 

Removing the permutation condition cannot decrease the number of equivalence classes, so there are infinitely many equivalence classes under phase--conjugation.

\end{proof}

\subsection{Proof of uncountably many PC-classes in dimensions $d = 4k$, where $k\in \mathbb{Z}_{>0}$} \label{d=k4^n}

\begin{lemma}
Let $d = 4k$ with $k\in \mathbb{Z}_{>0}$. Then the set of complex Hadamard matrices of order $d$
contains uncountably many distinct equivalence classes under row/column phase multiplications and complex conjugation.
\end{lemma}

\begin{proof}
We say $H \sim_{\mathrm{PC}} K$ if either $K=D_1HD_2$ or $K=D_1 H^\ast D_2$ for some
diagonal unitaries $D_1,D_2$.

\medskip
\noindent{\bf Step 1: a PC-invariant.}
For any matrix $H=(h_{ij})$ with nonzero entries, define the cross-ratio
\[
\chi(H):=\frac{h_{1,2}\,h_{2,3}}{h_{1,3}\,h_{2,2}}.
\]
If $H' = D_1 H D_2$, then $h'_{ij}=a_i\,h_{ij}\,b_j$ and the diagonal phases cancel in the
ratio, hence $\chi(H')=\chi(H)$. If $H'= H^\ast$, then $\chi(H')= \chi(H)^\ast$.
Therefore, the unordered pair $\{\chi(H), \chi(H)^\ast\}$ is invariant under $\sim_{\mathrm{PC}}$.

\medskip
\noindent{\bf Step 2: the $4\times4$ family.}
For the family $U(\alpha)$ in \eqref{eq:dim4_ub}, a direct computation gives
\[
\chi(U(\alpha))
=\frac{U(\alpha)_{1,2}\,U(\alpha)_{2,3}}{U(\alpha)_{1,3}\,U(\alpha)_{2,2}}
=\frac{\left(\tfrac12\right)\left(-\tfrac12\right)}{\left(\tfrac12\right)\left(\tfrac{i e^{i\alpha}}2\right)}
= i e^{-i\alpha}.
\]
Hence, if $U(\alpha)\sim_{\mathrm{PC}} U(\beta)$ then either
$i e^{-i\alpha}= i e^{-i\beta}$ or $i e^{-i\alpha}= (i e^{-i\beta})^\ast=-i e^{i\beta}$.
Equivalently,
\[
\alpha\equiv \beta \ (\mathrm{mod}\ 2\pi)\qquad\text{or}\qquad \alpha\equiv \pi-\beta \ (\mathrm{mod}\ 2\pi).
\]
Restricting to $\alpha,\beta\in[0,\pi/2)$ forces $\alpha=\beta$. Thus the set
$\{U(\alpha):\alpha\in[0,\pi/2)\}$ contains uncountably many PC-inequivalent matrices.

\medskip
\noindent{\bf Step 3: multiply by a Fourier matrix to reach $d=4k$.}
Let $F_k$ be the $k\times k$ Fourier complex Hadamard matrix, and define
\[
V_{k}(\alpha):=F_k\otimes U(\alpha).
\]
Then $V_{k}(\alpha)$ is a complex Hadamard matrix of order $4k$.
Now pick the same four positions in the $U(\alpha)$ part while keeping the Fourier indices
fixed (e.g.\ using the first row/column block). The Fourier factors cancel in the cross-ratio,
so
\[
\chi\!\left(V_{k}(\alpha)\right)=\chi\!\left(U(\alpha)\right)= i e^{-i\alpha}.
\]
Thus $V_{k}(\alpha)\sim_{\mathrm{PC}} V_{k}(\beta)$ with $\alpha,\beta\in[0,\pi/2)$ implies
$\alpha=\beta$. Since $[0,\pi/2)$ is infinite (indeed uncountable), there are uncountably many
PC-inequivalent complex Hadamard matrices in dimension $d=4k$.
\end{proof}

\subsection{Inequivalence of the $\sim_{PP}$ and $\sim_{PC}$ classifications} \label{inequivalence}

\begin{lemma}
The equivalence relations $\sim_{PP}$ (phase and permutation) and $\sim_{PC}$ (phase and complex conjugation)
induce inequivalent classifications of complex Hadamard matrices. In particular, neither classification refines the other in general.
\end{lemma}

\begin{proof}
We exhibit the two required counterexamples in different dimensions.

\smallskip
\noindent\textbf{(i) $\sim_{PP}$ does not imply $\sim_{PC}$.}
In dimension $d=5$, Remark~\ref{rem:single class} implies that there is a unique $\sim_{PP}$-class.
On the other hand, by the argument of Appendix~\ref{d=5} there are $72$ distinct $\sim_{PC}$-classes.
Hence there exist $H\neq K$ with $H\sim_{PP}K$ but $H\not\sim_{PC}K$.

\smallskip
\noindent\textbf{(ii) $\sim_{PC}$ does not imply $\sim_{PP}$.}
In dimension $d=4$, for the family $U(\alpha)$ in \eqref{eq:dim4_ub} one checks directly that
$\overline{U(\alpha)}=U(\pi-\alpha)$, and therefore $U(\alpha)\sim_{PC}U(\pi-\alpha)$.
However, under $\sim_{PP}$ the inequivalent classes of this family are parametrized by $\alpha\in[0,\pi)$
with $U(\alpha)\sim_{PP}U(\alpha+\pi)$ \cite{Tadej}, so for any $\alpha\in[0,\pi/2)$ we have
$U(\alpha)\not\sim_{PP}U(\pi-\alpha)$.

\smallskip
Together, (i) and (ii) show that $\sim_{PP}$ and $\sim_{PC}$ are inequivalent notions of classification.
\end{proof}

\section{Fourier reduction of the conventional Bell measurement} \label{app:bell-to-chm}

Here we make precise the remark that conventional Bell-basis measurements are maximally entangled measurements but need not be complex Hadamard measurements in the Schmidt-aligned basis, and show that a single fixed local rotation removes this gap. All objects below are as defined in Appendix~\ref{app:pc_class_num}: the Fourier unitary $F_d$, the Weyl operators $X,Z$ of Eqs.~\eqref{eq:Weyl_shift}--\eqref{eq:Weyl_phase}, and the covariance $F_d X^s = Z^s F_d$ established in the proof of Remark~\ref{rem:perm_rev_Fourier}.
Equivalently, the rotated basis can be implemented with a conventional Bell analyzer by applying $F_d^\dagger\otimes\I$ to the incoming measured systems before the analyzer.

\begin{theorem}[Fourier reduction of the conventional Bell measurement]
\label{thm:bell-to-chm}
Let the \emph{conventional generalized Bell measurement} be the maximally entangled measurement whose vectors $\ket{\Gamma_{mn}} = (E_{mn}^\ast\otimes\I)\ket{\Phi}$ are generated by the operator basis
\[
\mathcal{O}(M_\mathsf{N})
=\Big\{\,E^{\ast}_{mn}=\tfrac{1}{\sqrt d}\,X^{m}Z^{n}\,\Big\}_{m,n=0}^{d-1}.
\]
Applying the single, outcome-independent local unitary $F_d\otimes\I$ on $\mathcal{H}_\mathsf{N_A}\otimes\mathcal{H}_\mathsf{N_B}$ maps this measurement onto the one with operator basis
\[
\tilde E^{\ast}_{mn}=\tfrac{1}{\sqrt d}\,Z^{m}F_dZ^{n},
\qquad
H_{mn}:=\sqrt d\,\tilde E^{\ast}_{mn}=Z^{m}F_dZ^{n}\in\mathbb{H}_d .
\]
This rotated measurement is a complex Hadamard measurement whose operators all belong to the single phase--conjugation class $[F_d]_{\mathrm{PC}}$; it is therefore $G$-concurrence optimal and universally LU-deterministic.
\end{theorem}

\begin{proof}
Since $\sqrt d\,E^{\ast}_{mn}=X^{m}Z^{n}$ is unitary, $E^{\ast}_{mn}\in\mathcal{S}_\mathrm{MEM}$ and the conventional measurement is $G$-concurrence optimal by Lemma~\ref{G-factorization}. However,
\[
\braket{j|X^{m}Z^{n}|k}=\omega^{nk}\,\delta_{j,\,k+m},
\qquad\text{so}\qquad
\bigl|\braket{j|E^{\ast}_{mn}|k}\bigr|\in\Big\{0,\tfrac{1}{\sqrt d}\Big\},
\]
which is not the constant value $1/d$; hence $E^{\ast}_{mn}\notin\mathcal{S}_\mathrm{UBM}$ and the bare Bell measurement is not a complex Hadamard measurement in the computational (Schmidt) basis.

Because $\ket{\Gamma_{mn}}=(E^{\ast}_{mn}\otimes\I)\ket{\Phi}$ and $(F_d\otimes\I)(E^{\ast}_{mn}\otimes\I)\ket{\Phi}=(F_d E^{\ast}_{mn}\otimes\I)\ket{\Phi}$, the rotation acts on the operator basis as $E^{\ast}_{mn}\mapsto\tilde E^{\ast}_{mn}=F_d E^{\ast}_{mn}=\tfrac{1}{\sqrt d}F_d X^{m}Z^{n}$. Using the covariance $F_d X^{m}=Z^{m}F_d$,
\[
\tilde E^{\ast}_{mn}=\tfrac{1}{\sqrt d}\,Z^{m}F_dZ^{n}.
\]
Entrywise, with $Z^{m}$ scaling row $j$ by $\omega^{mj}$ and $Z^{n}$ scaling column $k$ by $\omega^{nk}$,
\[
\braket{j|\tilde E^{\ast}_{mn}|k}
=\tfrac{1}{\sqrt d}\,\omega^{mj}\,(F_d)_{jk}\,\omega^{nk}
=\tfrac{1}{d}\,\omega^{\,mj+jk+nk},
\qquad\text{so}\qquad
\bigl|\braket{j|\tilde E^{\ast}_{mn}|k}\bigr|=\tfrac{1}{d},
\]
giving $\tilde E^{\ast}_{mn}\in\mathcal{S}_\mathrm{UBM}$. Moreover $\sqrt d\,\tilde E^{\ast}_{mn}=Z^{m}F_dZ^{n}$ is a product of unitaries, hence unitary, so $\tilde E^{\ast}_{mn}\in\mathcal{S}_\mathrm{MEM}$. Therefore
\[
\tilde E^{\ast}_{mn}\in\mathcal{S}_\mathrm{UBM}\cap\mathcal{S}_\mathrm{MEM}=\mathcal{S}_\mathrm{CHM},
\qquad
H_{mn}=Z^{m}F_dZ^{n}\in\mathbb{H}_d .
\]
Finally, $Z^{m}$ and $Z^{n}$ are diagonal unitaries, so $H_{mn}=D_L F_d D_R$ with $D_L=Z^{m}$ and $D_R=Z^{n}$; every outcome is thus PC-equivalent to $F_d$, and all $d^{2}$ operators lie in the single class $[F_d]_{\mathrm{PC}}$.

Since $F_d\otimes\I$ is unitary, $\{\ket{\tilde\Gamma_{mn}}\}$ is again an orthonormal basis, i.e.\ a valid projective measurement; its operators lie in $\mathcal{S}_\mathrm{CHM}$ and share one PC-class, so by Theorem~\ref{thm:phase-conjugation} it is universally LU-deterministic and by Lemma~\ref{G-factorization} it is $G$-concurrence optimal. The stated dimension dependence of $[F_d]_{\mathrm{PC}}$ follows from Theorem~\ref{thm:PC_class_count}.
\end{proof}

\section{Noise Robustness} \label{app:noise_robustness}

\subsubsection{Robustness to Depolarizing Noise}

We consider mixed inputs obtained by applying bipartite depolarizing noise to each link:
\begin{align}
\rho_{\mathsf{A N_A}} &= (1-p)\,\proj{\psi} + p\,\frac{\mathds{1}}{d^2},\qquad
\rho_{\mathsf{N_B B}} = (1-q)\,\proj{\phi} + q\,\frac{\mathds{1}}{d^2},
\label{eq:mixed_input}
\end{align}
where the ideal pure states are
\begin{align}
\ket{\psi} = (A\otimes \mathds{1})\ket{\Phi},\qquad
\ket{\phi} = (B\otimes \mathds{1})\ket{\Phi},
\end{align}
with $A,B$ diagonal, positive semidefinite, and normalized as $\|A\|_F=\|B\|_F=1$.
Accordingly, the reduced states are diagonal in the computational basis:
\begin{align}
\rho_\psi := \tr_{\mathsf{N_A}}[\proj{\psi}] = A^2,\qquad
\rho_\phi := \tr_{\mathsf{N_B}}[\proj{\phi}] = B^2.
\label{eq:reduced_states_diag}
\end{align}

The node performs a projective measurement $\{M_i\}_{i=1}^{d^2}$ on $\mathsf{N_A N_B}$:
\begin{align}
M_i = \mathds{1}_\mathsf{A}\otimes \proj{\Gamma_i}_{\mathsf{N_A N_B}}\otimes \mathds{1}_\mathsf{B},
\qquad
\ket{\Gamma_i} = (E_i^\ast\otimes \mathds{1})\ket{\Phi}_{\mathsf{N_A N_B}},
\end{align}
and we assume that $\{E_i^\ast\}_{i=1}^{d^2}$ is an operator basis of a complex Hadamard measurement.

\paragraph*{Diagonal PC-orbit assumption.}
For robustness, we impose the following \emph{sufficient} structural condition: all measurement operators 
lie in the same \emph{phase-multiplication orbit} (i.e., the non-conjugate part of a PC-class),
\begin{align}
E_i = D_L^{(i)}\,E_1\,D_R^{(i)} \qquad \text{for all $i$},
\label{eq:diag_orbit_assumption}
\end{align}
where $D_L^{(i)},D_R^{(i)}$ are diagonal unitaries in the computational basis.
(We discuss the conjugation branch $E_i=D_L^{(i)} E^\ast_1 D_R^{(i)}$ in Remark~\ref{rem:conjugation_obstruction} below.)

\paragraph*{Conditional output states.}
For outcome $i$, the unnormalized post-measurement state on $\mathsf{AB}$ is
\begin{align}
\tilde{\eta}_{i,\mathsf{AB}}
:= \tr_{\mathsf{N_A N_B}}\!\left[ M_i\,
\bigl(\rho_{\mathsf{A N_A}}\otimes \rho_{\mathsf{N_B B}}\bigr) \right].
\end{align}
A direct expansion using Eq.~\eqref{eq:mixed_input} yields
\begin{align}
\tilde{\eta}_{i,\mathsf{AB}}
= \frac{1}{d^2}\Bigl[
(1-p)(1-q)\,\proj{\eta_i}
+ (1-p)q \,\rho_\psi \otimes \frac{\mathds{1}}{d}
+ p(1-q)\,\frac{\mathds{1}}{d} \otimes \rho_\phi
+ pq\,\frac{\mathds{1}}{d}\otimes \frac{\mathds{1}}{d}
\Bigr],
\label{eq:mixed_output_unnormalized}
\end{align}
where the pure-output vectors are
\begin{align}
\ket{\eta_i} = \bigl(d\,A E_i B \otimes \mathds{1}\bigr)\ket{\Phi}.
\label{eq:eta_i_def_noise_app}
\end{align}
Taking the trace of Eq.~\eqref{eq:mixed_output_unnormalized} and using the normalization of its four component states gives $p_i:=\tr(\tilde{\eta}_{i,\mathsf{AB}})=1/d^2$.
Hence the normalized conditional output state is
\begin{align}
\eta_{i,\mathsf{AB}} = d^2\,\tilde{\eta}_{i,\mathsf{AB}},
\end{align}
i.e., the bracketed expression in Eq.~\eqref{eq:mixed_output_unnormalized}.

\paragraph*{LU-determinism under depolarizing noise (sufficient condition).}
By the phase-multiplication assumption~\eqref{eq:diag_orbit_assumption} and the diagonality of $A,B$, we have
\begin{align}
A E_i B = D_L^{(i)}\,(A E_1 B)\,D_R^{(i)}.
\end{align}
Using $(U\otimes V)(C\otimes \mathds{1})\ket{\Phi}=(U C V^T\otimes \mathds{1})\ket{\Phi}$ and the fact that diagonal matrices satisfy $D^T=D$,
it follows that
\begin{align}
\ket{\eta_i} = (D_L^{(i)}\otimes D_R^{(i)})\,\ket{\eta_1}.
\label{eq:pure_corrections_diagonal}
\end{align}
Since $\rho_\psi$ and $\rho_\phi$ are diagonal in the same basis by Eq.~\eqref{eq:reduced_states_diag},
the same diagonal unitaries commute with the noise terms:
\begin{align}
D_L^{(i)}\,\rho_\psi\,D_L^{(i)\dagger}=\rho_\psi,\qquad
D_R^{(i)}\,\rho_\phi\,D_R^{(i)\dagger}=\rho_\phi.
\end{align}
Combining this with Eq.~\eqref{eq:mixed_output_unnormalized} shows that the mixed outputs are mutually LU-equivalent:
for all $i$,
\begin{align}
\eta_{i,\mathsf{AB}}
= (D_L^{(i)}\otimes D_R^{(i)})\,\eta_{1,\mathsf{AB}}\,(D_L^{(i)}\otimes D_R^{(i)})^\dagger.
\end{align}
Therefore, the universal LU-determinism of the swapping measurement is robust to bipartite depolarizing noise
under the sufficient condition~\eqref{eq:diag_orbit_assumption}.

\begin{remark}[Obstruction from the conjugation branch]\label{rem:conjugation_obstruction}
If some outcomes satisfy $E_i = D_L^{(i)} E^\ast_1 D_R^{(i)}$ with $E_1\not\sim D_L E^\ast_1 D_R$ (i.e., $E_1$ is not
self-conjugate up to diagonal phases), then the local unitaries mapping $\ket{\eta_1}\mapsto \ket{\eta_i}$ in the pure case
need not be diagonal. In that situation, for generic inputs with nondegenerate $\rho_\psi$ and $\rho_\phi$,
such non-diagonal corrections do not preserve the noise terms $\rho_\psi\otimes \mathds{1}/d$ and $\mathds{1}/d\otimes \rho_\phi$,
and mixed-state LU-determinism can fail. 
\end{remark}

\subsubsection{Robustness to General Noise} \label{app:general_robustness}
Alice and the node share a bipartite pure state

In experimental implementations, the input states may deviate from the ideal pure states due to noise or imperfections. We characterize the robustness of the protocol by analyzing the average distance between the actual output states and the ideal target states.

Consider a noisy input state $\rho$ acting on $\mathcal{H}_\mathsf{A} \otimes \mathcal{H}_\mathsf{N_A} \otimes \mathcal{H}_\mathsf{N_B} \otimes \mathcal{H}_\mathsf{B}$, such that
\begin{align}
\rho = (1-\epsilon) \ket{\psi} \bra{\psi} \otimes \ket{\phi} \bra{\phi}  + \epsilon \sigma_{\mathrm{noise}},
\end{align}
where $\ket{\psi} \bra{\psi} \otimes \ket{\phi} \bra{\phi}$ corresponds to the ideal input and $\sigma_{\mathrm{noise}}$ represents the normalized residual noise.

Let the node perform the entanglement swapping measurement described in the protocol. This defines a quantum instrument acting on the input state. Due to the linearity of the map, the unnormalized conditional output for outcome $i$ is
\begin{align}
\tilde{\rho}_i = \frac{(1-\epsilon)}{d^2} \proj{\eta_i} + \epsilon \tilde{\sigma}_i,
\end{align}
where $\tilde{\sigma}_i$ is the unnormalized noise contribution. The probability of obtaining outcome $i$ is $p_i = \tr(\tilde{\rho}_i) = (1-\epsilon)d^{-2} + \epsilon q_i$, where $q_i = \tr(\tilde{\sigma}_i)$.

The normalized output state is $\rho_i = \tilde{\rho}_i / p_i$. We quantify the performance via the \emph{average trace distance} $\overline{D}$, defined as the trace distance between the actual and ideal outputs averaged over the measurement statistics:
\begin{align}
\overline{D} = \sum_i \frac{p_i}{2} \| \rho_i - \proj{\eta_i} \|_1.
\end{align}

\begin{lemma}
For any input state of the form $\rho=(1-\epsilon)\proj{\psi}\otimes\proj{\phi}+\epsilon\sigma_{\mathrm{noise}}$, with $0\leq\epsilon<1$, the average trace distance between the generated outputs and the ideal LU-deterministic outputs satisfies
\begin{align}
\overline{D} \leq \epsilon.
\end{align}
\end{lemma}

\begin{proof}
We start with 
\begin{align}
\rho_i - \proj{\eta_i} &= \frac{1}{p_i} \left[ \frac{1-\epsilon}{d^2}\proj{\eta_i} + \epsilon \tilde{\sigma}_i - \left( \frac{1-\epsilon}{d^2} + \epsilon q_i \right) \proj{\eta_i} \right] \nonumber \\
&= \frac{\epsilon}{p_i} (\tilde{\sigma}_i - q_i \proj{\eta_i}).
\end{align}
Inserting this expression into the definition of $\overline{D}$, the probability $p_i$ cancels out:
\begin{align}
\overline{D} &= \frac{\epsilon}{2} \sum_i \| \tilde{\sigma}_i - q_i \proj{\eta_i} \|_1.
\end{align}
Using the triangle inequality and observing that $\|\tilde{\sigma}_i\|_1 = q_i$ and $\|\proj{\eta_i}\|_1 = 1$:
\begin{align}
\| \tilde{\sigma}_i - q_i \proj{\eta_i} \|_1 \leq \|\tilde{\sigma}_i\|_1 + q_i \|\proj{\eta_i}\|_1 = 2q_i.
\end{align}
Finally, since $\sum_{i=1}^{d^2} p_i = \sum_{i=1}^{d^2} (\frac{1-\epsilon}{d^2} + \epsilon q_i )= 1 $ we have $\sum_{i=1}^{d^2} q_i = 1$, and then
\begin{align}
\overline{D} \leq \frac{\epsilon}{2} \sum_i 2q_i = \epsilon.
\end{align}
\end{proof}

\section{Entanglement Distribution along a Network Path} \label{app:general_network}
\subsubsection{Settings and definitions}

In this section, we consider a chain
\[
\mathsf{A}\!-\!\mathsf{N}_1\!-\!\cdots\!-\!\mathsf{N}_L\!-\!\mathsf{B},
\]
and prove that the corrected end-to-end state is \emph{order-independent} for $d\in\{2,3\}$.

Consider a single swapping node holding subsystems $\mathsf{N,N'}$ that are entangled with terminals $\mathsf{A,B}$ via input states
$\ket{\psi(J)}_\mathsf{AN} = (J \otimes \I)\ket{\Phi}$ and $\ket{\psi(K)}_\mathsf{N'B} = (K \otimes \I)\ket{\Phi}$, where $J,K$ are positive, diagonal, and normalized: $\Vert J \Vert_\mathrm{F}= \Vert K \Vert_\mathrm{F}=1$.
Let the swapping node perform a complex Hadamard measurement on $\mathsf{NN'}$ generated by coefficient matrices PC-equivalent to the Fourier unitary $F$.

As established earlier, the corrected output in the computational Schmidt basis is independent of the measurement outcome up to local unitaries and can be written as $\ket{\psi(L)}_\mathsf{AB}$ with diagonal $L$. Note that $L$ is given by
\[
L = \sqrt d \operatorname{diag}(\sv(JFK)),
\]
where $\sv$ means the non-increasingly ordered singular value vector, $F$ is the Fourier matrix, and $\sqrt d$ is multiplied for normalization $\tr[L\dg L] = 1$.
This Fourier-based composition rule coincides with the deterministic series rule of Ref.~\cite{MengEtAl2023}.

\begin{definition}[Fusion operation $\star$]
We then define the \emph{fusion} of $J$ and $K$ by
\begin{align}
L = J \star K = \sqrt d \operatorname{diag}(\sv(JFK)).
\end{align}
\end{definition}
Note that the G-concurrence of the states is multiplicative:
\begin{align}
d \abs{\det L }^{2/d} = d \abs{\det J }^{2/d} \cdot d \abs{\det K }^{2/d}
\end{align}

By construction, $\star$ is a map that takes pairs of diagonal coefficient matrices to a diagonal coefficient matrix.

\subsubsection{Two-swapping nodes}
We start with the simplest scenario with two swapping nodes.
Consider a chain $\mathsf{A}\!-\!\mathsf{N}_1\!-\!\mathsf{N}_2\!-\!\mathsf{B}$ with input pure states
\begin{align}
\ket a_\mathsf{AN_1} &= (A \otimes \I) \ket{\Phi}, \nonumber \\
\ket b_\mathsf{N_1' N_2} &= (B \otimes \I) \ket{\Phi}, \nonumber \\
\ket c_\mathsf{N_2'B} &= (C \otimes \I) \ket{\Phi},
\end{align}
where $A,B,C$ are diagonal coefficient matrices.

\begin{theorem}[Associativity for two swapping nodes in low dimension]\label{thm:two-node}
Let $d\in\{2,3\}$ and $\star$ be the fusion operation. Then
\[
(A\star B)\star C \;=\; A\star(B\star C).
\]
Equivalently, after correction to computational Schmidt form, the final diagonal coefficient matrix is independent of whether $\mathsf{N}_1$ or $\mathsf{N}_2$ swaps first.
\end{theorem}
 
For $d=2$, the theorem follows immediately from normalization and Lemma~\ref{G-factorization}, because concurrence fully determines the Schmidt coefficients of a pure qubit state.
We now prove the $d=3$ case.
\begin{proof}
Let $AFB=U_1D_1V_1$ and $BFC=U_2D_2V_2$ be singular-value decompositions, and define $M_1=D_1FC$ and $M_2=AFD_2$.
It suffices to show that $M_1$ and $M_2$ have the same multiset of singular values.

Define the Hermitian matrices
\begin{align}
H_1 &= M_1^\dagger M_1=C^\dagger F^\dagger D_1^2FC,\\
H_2 &= M_2^\dagger M_2=D_2^\dagger F^\dagger A^2FD_2.
\end{align}
For $3\times3$ Hermitian matrices, equality of $\tr(H_i)$, $\det(H_i)$, and $\tr(H_i^2)$ implies equality of their characteristic polynomials.
The first two invariants already coincide:
\begin{align}
\tr(H_1)=\tr(H_2)=\frac{1}{9},
\qquad
\det(H_1)=\det(H_2)=(\det A)^2(\det B)^2(\det C)^2.
\end{align}
It remains to prove that $\tr(H_1^2)=\tr(H_2^2)$.
We use the Weyl shift and phase operators $X$ and $Z$ defined in Eqs.~\eqref{eq:Weyl_shift} and \eqref{eq:Weyl_phase} with $d=3$.
For $d=3$, a diagonal matrix $G=\operatorname{diag}(g_0,g_1,g_2)$ can be decomposed as
\begin{align}
G &= \sum_{t=0}^{2} \tilde g_t Z^t,
\qquad
\tilde g_t
= \frac{1}{3}\tr[Z^{-t}G]
= \frac{1}{3}\sum_{k=0}^{2}\omega^{-kt}g_k,
\end{align}
implying $F\dg G F=\sum_{t=0}^{2}\tilde g_t X^t$.

In the following, all Weyl indices and Kronecker deltas are understood modulo $3$.
Define
\begin{align}
J &= F\dg A^2 F = \sum_t \alpha_t X^t, \\
K &= F C^2 F\dg = \sum_t \gamma_t X^t.
\end{align}
and note that $\alpha_0 = \gamma_0 = \frac{1}{3}$, $\alpha_{-t}=\overline{\alpha_t}$, and $\gamma_{-t}=\overline{\gamma_t}$. 
Additionally, define $L:=D_1^2$ and $N:=D_2^2$, and note that
\begin{align}
\tr(L) &= \tr[D_1^2] = \tr[V_1\dg D_1\dg U_1\dg U_1 D_1 V_1] = \tr[F\dg A^2 F B^2] \nonumber \\
&= \sum_{i} \braket{i|F\dg A^2 F B^2|i} = \sum_i b_i^2 \sum_{jk} \braket{i|F\dg|j}\braket{j|A^2|k}\braket{k|F|i} \nonumber \\
&= \sum_{ij} b_i^2 a_j^2 |\braket{i|F|j}|^2 = \frac{1}{3} \tr[A^2] \tr[B^2] = \frac{1}{3}, \\
\tr(N) &= \frac{1}{3}\tr[B^2]\tr[C^2] = \frac{1}{3}.
\end{align}
Define
\begin{align}
S_t(G) = \sum_{k=0}^{2} g_k g_{k+t} ~~\text{for}~~ G = \operatorname{diag}(g_0, \ldots, g_{2}).
\end{align}
Then we calculate
\begin{align}
\tr[H_1^2] &= \tr[LKLK] \\
&= \sum_{t,u} \gamma_t \gamma_u \tr[L X^t L X^u] \\
&= \sum_{t,u} \gamma_t \gamma_u \delta_{t+u,0} S_t(L) \\
&= \sum_t |\gamma_t|^2 S_t(L).
\end{align}
Note that for $d=3$,
\begin{align}
S_1(L) &= S_2(L) = \frac{1}{2} ( (\tr L)^2 - \tr (L^2)), \\
S_0(L) &= \tr(L^2) \\
\tr[H_1^2] &= (|\gamma_0|^2-|\gamma_1|^2) \tr(L^2) + \frac{1}{9} |\gamma_1|^2
\end{align}
\begin{align}
\tr(L^2) &= \tr[(B F\dg A^2 F B)^2] = \tr[ (\sum_r \alpha_r B X^r B)^2 ] \\
&= \sum_{rs} \alpha_r \alpha_s \tr[B^2 X^r B^2 X^s] \\
&= \sum_{rs} \alpha_r \alpha_s \delta_{r+s,0} S_r(B^2) = \sum_{r} |\alpha_r|^2 S_r(B^2) \\
&= |\alpha_0|^2 S_0(B^2) + 2|\alpha_1|^2 S_1(B^2)
\end{align}
Therefore,
\begin{align}
\tr[H_1^2] &= (|\gamma_0|^2-|\gamma_1|^2) \tr(L^2) + \frac{1}{9} |\gamma_1|^2 \\
&= \frac{1}{9} |\gamma_1|^2  + (\frac{1}{9} - |\gamma_1|^2) (\frac{1}{9} S_0(B^2) + 2|\alpha_1|^2 S_1(B^2)) \label{eq:trH1sq}
\end{align}
Similarly,
\begin{align}
\tr[H_2^2] &= \tr[NJNJ] \\
&= \sum_{t,u} \alpha_t \alpha_u \tr[N X^t N X^u] \\
&= \sum_{t,u} \alpha_t \alpha_u \delta_{t+u,0} S_t(N) \\
&= \sum_t |\alpha_t|^2 S_t(N).
\end{align}
For $d=3$,
\begin{align}
S_1(N) &= S_2(N) = \frac{1}{2} ( (\tr N)^2 - \tr (N^2)), \\
S_0(N) &= \tr(N^2).
\end{align}

\begin{align}
\tr(N^2) &= \tr[(B F C^2 F\dg B)^2] = \tr[ (\sum_r \gamma_r B X^r B)^2 ] \\
&= \sum_{rs} \gamma_r \gamma_s \tr[B^2 X^r B^2 X^s] \\
&= \sum_{rs} \gamma_r \gamma_s \delta_{r+s,0} S_r(B^2) = \sum_{r} |\gamma_r|^2 S_r(B^2) \\
&= |\gamma_0|^2 S_0(B^2) + 2|\gamma_1|^2 S_1(B^2).
\end{align}
Thus,
\begin{align}
\tr[H_2^2] &= (|\alpha_0|^2-|\alpha_1|^2) \tr(N^2) + \frac{1}{9} |\alpha_1|^2 \\
&= \frac{1}{9} |\alpha_1|^2  + (\frac{1}{9} - |\alpha_1|^2) (\frac{1}{9} S_0(B^2) + 2|\gamma_1|^2 S_1(B^2))  \label{eq:trH2sq}
\end{align}

Subtracting Eq.~\eqref{eq:trH2sq} from Eq.~\eqref{eq:trH1sq} yields
\begin{align}
\tr[H_1^2]  - \tr[H_2^2] = \frac{1}{9}(|\alpha_1|^2-|\gamma_1|^2)(S_0(B^2) + 2S_1(B^2)-1) = 0,
\end{align}
because
\begin{align}
S_0(B^2) + 2S_1(B^2) = S_0(B^2) + S_1(B^2) + S_2(B^2) = \tr[B^2]^2 = 1.
\end{align}

Combining all results, we have
\[
\tr H_1 = \tr H_2, ~~ \det H_1 = \det H_2, ~~ \tr H_1^2 = \tr H_2^2,
\]
implying $H_1 = M_1\dg M_1$ and $H_2 = M_2\dg M_2$ have the same multiset of eigenvalues, thus $M_1$ and $M_2$ share the singular value vector.
\end{proof}

\subsubsection{General linear chain}
Now we consider a general $L$-node chain.
In a chain $\mathsf{A}\!-\!\mathsf{N}_1\!-\!\cdots\!-\!\mathsf{N}_L\!-\!\mathsf{B}$ with input states
$\ket{\psi(A_0)}_{\mathsf{A}\mathsf{N}_1}, \ket{\psi(A_1)}_{\mathsf{N}_1'\mathsf{N}_2}, \ldots, \ket{\psi(A_L)}_{\mathsf{N}_L'\mathsf{B}}$,
each full parenthesization $\mathbf s$ of the word $0\,1\,\cdots\,L$ (with $L$ binary fusions) specifies an evaluation order for the iterated fusion
\[
  F_{\mathbf s} \;=\; A_0 \star A_1 \star \cdots \star A_L
\]
computed according to $\mathbf s$.
(Equivalently, $F_{\mathbf s}$ is defined recursively by replacing each innermost pair $(A_i,A_{i+1})$ with $A_i\star A_{i+1}$ until a single diagonal remains.)

\begin{corollary}[Order independence for $d\in\{2,3\}$]
\label{cor:order-independence}
Let $d\in\{2,3\}$ and consider the $L$-node chain above.
For any two full parenthesizations $\mathbf s,\mathbf s'$ of $0\,1\,\cdots\,L$,
\[
F_{\mathbf s} \;=\; F_{\mathbf s'}.
\]
Thus all swapping strategies yield the same final corrected diagonal coefficient matrix (and hence the same LU class of the shared output state between $\mathsf{A}$ and $\mathsf{B}$).
\end{corollary}
\begin{proof}   
For $n\ge 1$ and diagonal matrices $A_0,\ldots,A_{n-1}$, define the left-associated evaluation
\[
\mathrm{L}(A_0)=A_0,\qquad
\mathrm{L}(A_0,\ldots,A_{n}) := \mathrm{L}(A_0,\ldots,A_{n-1})\star A_{n} \quad(n\ge 1).
\]

\emph{Key lemma. }
For any diagonal $X$ and any sequence $B_1,\ldots,B_m$ with $m\ge 1$,
\begin{equation}\label{eq:key}
X \star \mathrm{L}(B_1,\ldots,B_m) \;=\; \mathrm{L}(X,B_1,\ldots,B_m).
\end{equation}
\emph{Proof of the lemma. }By induction on $m$.
For $m=1$, both sides equal $X\star B_1$.
Assume \eqref{eq:key} for $m-1\ge 1$.
Write $\mathrm{L}(B_1,\ldots,B_m)=\mathrm{L}(B_1,\ldots,B_{m-1})\star B_m$.
Then, using Theorem~\ref{thm:two-node} (associativity),
\[
X\star \mathrm{L}(B_1,\ldots,B_m)
= X\star\big(\mathrm{L}(B_1,\ldots,B_{m-1})\star B_m\big)
= \big(X\star \mathrm{L}(B_1,\ldots,B_{m-1})\big)\star B_m
= \mathrm{L}(X,B_1,\ldots,B_m),
\]
where the last equality uses the induction hypothesis. $\square$

We prove by induction on $n=L{+}1$ that for every full parenthesization $\mathbf s$ of $A_0,\ldots,A_{n-1}$,
\begin{equation}\label{eq:goal}
F_{\mathbf s} \;=\; \mathrm{L}(A_0,\ldots,A_{n-1}).
\end{equation}
The claim is trivial for $n=1,2$.
Assume it holds for all lengths $<n$ and consider any full parenthesization $\mathbf s$ of length $n$.
Let the top split (the last fusion before the final output) of $\mathbf s$ be $(\mathbf s_{\mathrm L}\star \mathbf s_{\mathrm R})$, where
$\mathbf s_{\mathrm L}$ (resp. $\mathbf s_{\mathrm R}$) parenthesizes the contiguous block
$A_0,\ldots,A_k$ (resp. $A_{k+1},\ldots,A_{n-1}$) for some $0\le k\le n-2$.
By the induction hypothesis,
\[
F_{\mathbf s_{\mathrm L}}=\mathrm{L}(A_0,\ldots,A_k),\qquad
F_{\mathbf s_{\mathrm R}}=\mathrm{L}(A_{k+1},\ldots,A_{n-1}).
\]
Therefore
\[
F_{\mathbf s}
= F_{\mathbf s_{\mathrm L}}\star F_{\mathbf s_{\mathrm R}}
= \mathrm{L}(A_0,\ldots,A_k)\star \mathrm{L}(A_{k+1},\ldots,A_{n-1})
= \mathrm{L}(A_0,\ldots,A_{n-1}),
\]
where the last equality is exactly the lemma \eqref{eq:key} with $X=\mathrm{L}(A_0,\ldots,A_k)$ and
$(B_1,\ldots,B_m)=(A_{k+1},\ldots,A_{n-1})$.
Thus \eqref{eq:goal} holds for length $n$, completing the induction.

Consequently, $F_{\mathbf s}$ is independent of $\mathbf s$, and every swapping strategy yields the same final corrected diagonal coefficient matrix (hence the same LU class between $\mathsf A$ and $\mathsf B$).
\end{proof}

\begin{remark}
Associativity need not hold beyond $d=3$. The following $d=4$ example provides a counterexample:
\begin{align*}
A = \frac{1}{\sqrt{164}} \operatorname{diag}(9,9,1,1), \\
B = \frac{1}{\sqrt{252}} \operatorname{diag}(9,9,9,3), \\
C = \frac{1}{\sqrt{115}} \operatorname{diag}(8,5,5,1).
\end{align*}
Then
\begin{align*}
(A \star B) \star C &= \operatorname{diag}(0.856,0.512,0.070,0.013) \\
\neq
A \star (B \star C) &= \operatorname{diag}(0.879,0.471,0.075,0.013)
\end{align*}
\end{remark}

\end{document}